\documentclass[%
 reprint,
superscriptaddress,
nobibnotes,
 amsmath,amssymb,
 aps,
 prd,
floatfix,
]{revtex4-2}

\usepackage{booktabs}
\usepackage{graphicx}% Include figure files
\usepackage{dcolumn}% Align table columns on decimal point
\usepackage{bm}% bold math
\usepackage{verbatim}
\usepackage{placeins}
\usepackage{afterpage}
\usepackage{caption}
\usepackage{hyperref}% add hypertext capabilities
\usepackage{todonotes}

\newcommand{\mnras}{Mon.~Not.~Roy.~Astron.~Soc.}

\newcommand{\jcap}{JCAP}
\newcommand{\pasj}{Publ.~Astron.~Soc.~Japan}

\newcommand{\aap}{Astronomy~\&~Astrophysics}

\newcommand{\araa}{Annu.~Rev.~Astron.~Astrophys.}
\newcommand{\ssr}{Space Science Reviews}

\DeclareUnicodeCharacter{200B}{}

\begin{document}

\preprint{APS/123-QED}

\title{Splashback radius and the mass accretion rate of RASS MCMF galaxy clusters}% Force line breaks with \\

\author{Jitendra Joshi}
\affiliation{Inter University Centre for Astronomy and Astrophysics, Ganeshkhind, Pune 411007, India}
% \altaffiliation{Physics Department, XYZ University.}%Lines break automatically or can be forced with \\
\author{Divya Rana}%
\affiliation{Leiden Observatory, Leiden University, PO Box 9513, NL-2300 RA Leiden, The Netherlands}
\affiliation{Inter University Centre for Astronomy and Astrophysics, Ganeshkhind, Pune 411007, India}
\author{Surhud More}
\affiliation{Inter University Centre for Astronomy and Astrophysics, Ganeshkhind, Pune 411007, India}
\affiliation{Kavli Institute for the Physics and Mathematics of the Universe (WPI), University of Tokyo, 5-1-5, Kashiwanoha, 2778583, Japan}
\author{Matthias Klein}
\affiliation{Universitaets-Sternwarte Muenchen, Scheinerstrasse 1, 81679 München, Germany}
% \affiliation{%

\date{\today}% It is always \today, today,

\begin{abstract}
We present measurements of the radial profile of mass and galaxy number density around X-ray selected ROSAT All Sky Survey-Multi-Component Matched Filter galaxy clusters using Year 3 data from the Dark Energy Survey. We measure the projected cross-correlation signal of the RedMaGiC ``high density" galaxies around an approximately volume-limited sample of 255 galaxy clusters at a median redshift of $z=0.4$ and an X-ray luminosity $L_X > 10^{44} \,\text{ergs} \, \text{s}^{-1} \, \text{h}^{-2}$. This cross-correlation signal measured with a signal-to-noise ratio of 16.41 allows us to infer a 3D number density profile which shows a significant steepening at the edges of these galaxy clusters, namely the splashback radius of $r_{sp}$ /$h^{-1} \mathrm{M_{\odot}} = 2.19^{+0.50}_{-0.43}$. We present the dependence of the splashback radius value over a range of absolute galaxy magnitude cuts to look for any evidence of dynamical friction affecting these results. The weak lensing signal around our galaxy clusters measured with a signal-to-noise ratio of 32.19 allows us to infer a halo mass $\text{log} (M_{\rm 200m} / h^{-1} \text{Mpc}) = 14.68_{-0.04}^{+0.04}$. Comparison of the location of the splashback radius with the spherical overdensity boundary $r_{\rm 200m}$ shows consistency with the $\mathrm{\Lambda CDM}$ predictions. We present the first inference of the average mass accretion rate of galaxy clusters using our measurements of the splashback radius.
\begin{description}
\item[Usage]
Secondary publications and information retrieval purposes.
% You may use the \texttt{description} environment to structure your abstract;
% use the optional argument of the \verb+\item+ command to give the category of each item. 
\end{description}
\end{abstract}

\maketitle

%\tableofcontents

% %\section{Test}
% This is a test of \citet{Divya2023} with another \citep{Divya2023} and another example \cite{Divya2023}. \parencite{Divya2023}.
%\printbibliography

\section{\label{sec:level1}Introduction}
As the largest collapsed structures in the Universe, galaxy clusters offer a unique window into the growth of cosmic structure over time. According to the standard Lambda cold dark matter ($\mathrm{\Lambda}$CDM) model, galaxy cluster sized dark matter halos form through hierarchical gravitational collapse of high-density peaks in the initial matter distribution (see; \cite{Huss1999, Kravtsov2012, Walker2019,Vogelsberger2020}). The matter distribution and the size of these halos play an important role in understanding the structure formation in our Universe. The Navarro-Frenk-White (NFW) proﬁle \cite{NFW_profile} where the density drops as $r^{-1}$ closer to the center and gradually changes to $r^{-3}$ has been shown to describe well the inner regions of these dark matter halos. The NFW profile is self-similar, meaning that it can be described by a single functional form that depends on just two parameters: the characteristic density and scale radius. 

However, numerical simulations of collisionless dark matter have shown that in the outskirts of dark matter halos, especially on cluster scales, the logarithmic slopes of the density distribution can be significantly steeper than the expectations from the NFW profile value of $-3$ \cite{Diemer_Kravtsov_2014}. They showed that the density profiles can be better fitted with an alternative form (hereafter, DK14 profile) which is a combination of an Einasto profile that smoothly transitions to a power law profile on large scales. \citet{More2015} characterized the stacked mass profiles of haloes at different redshifts and stages of evolution. This has been studied and investigated in several further studies, e.g., \citet{Adhikari2014, Shi2016} explore the theoretical models that result in the profiles, and especially, the location of the steepest slope, also called the splashback radius. \citet{Diemer2017} and \citet{Mansfield2017} have further studied the characterization of the matter density profiles for individual halos in dark matter simulations.

Traditionally, the size of galaxy clusters is calculated using the location where the density of the halo is equal to $200$ times the critical or mean density of the Universe ($R_{\rm 200c}$ and $R_{\rm 200m}$). However, the location of the splashback radius, $r_{\rm sp}$, is a more physical choice for the  halo boundary \cite{More2015} that takes care of issues such as pseudo-evolution \cite{Diemer2013, Masato2019}, satellites outside the virial radius, $R_{\rm vir}$ \cite{Wetzel2014}, and infalling subhalos getting stripped far outside $R_{\rm vir}$ \cite{Behroozi2014}. The term splashback radius was coined by \citet{Adhikari2014} to indicate that the location marks the boundary of the multi-stream regime, where particles that have undergone a peri-centric passage through the halo, mix with infalling material at this location. Splashback radius has been found to be primarily dependent on the redshift, mass and accretion rate of the halo \cite{More2015, Diemer2017, Donnell2021}. 

These results have led to various studies on the splashback radius. Over the past decade, a rich body of simulation-based studies has advanced our understanding of the splashback feature (see e.g. \cite{Mansfield2017, Okumura2018, Fong2018, Mansfield2020, Sugiura2020, Diemer2020, Xhakaj2020, Deason2021, ONeil2021, Contigiani2021, Towler2024, Zhang2024}). The splashback feature has also emerged as a useful tool in cosmological analyses using different dark matter and dark energies theories \cite{Adhikari2018, Contigiani2019, Banerjee2020, Haggar2024} and as a test of galaxy evolution theories in galaxy clusters \cite{Dacunha2022}. 

The mass accretion rate of halos is important to understand the formation and evolution of galaxy clusters \cite{1993Lacey,2002Vandenbosch,2012Giocoli} and is closely linked to various halo properties, including spin \cite{2002Vitvitska}, shape \cite{2005Kasun,2006Allgood}, concentration  \cite{2002Wechsler,2009Zhao,2013Ludlow}, internal structure \cite{2004Gao,2005Vandenbosch,2012Power, 2013Ludlow}, and the input cosmology \cite{2002Vandenbosch,2015Correa}. In numerical simulations, these accretion rates are typically analyzed through halo merger trees constructed across different redshifts \cite{2008Genel,2012Kuhlen}. In contrast, observational studies lack such detailed merger histories for individual halos. However, the splashback radius given its sensitivity to the mass accretion rate \cite{Diemer_Kravtsov_2014, More2015} of dark matter halos, offers a promising, direct observational measure for these rates. 

The splashback radius is especially useful in the study of galaxy clusters, where the density drop at this scale is more prominent due to their typically high accretion rates. Galaxy clusters also suffer less from contamination of satellite halos which can be a significant problem for galaxy or galaxy group halos. Studies have shown that different selection methods yield different $r_{\rm sp}$ results because of the various systematics involved. For instance, studies have shown that optically selected clusters tend to yield smaller $r_{\rm sp}$ values than expected from numerical simulations, potentially due to biases in galaxy population representation and selection criteria \cite{More2016, Baxter2017, Chang2018, Sunayama2019, Murata2020, Giocoli2024}. On the other hand, SZ-selected clusters provide a more consistent estimate of $r_{sp}$, as they are less affected by optical selection biases and better capture the underlying dark matter halo properties \cite{zuricher2019, Shin2019, Shin2021}. These results show consistency with $\Lambda$CDM predictions. However, the substantial errors associated with these estimates do not entirely rule out the initial findings derived from optically selected clusters. Even after the use of more extensive SZ selected cluster catalogs, they find that the precision of the location of the splashback radius is still not competitive with that of optically selected galaxy clusters, primarily due to the sample size.

X-ray observations are particularly effective in identifying the hot gas within clusters, allowing for a clearer selection of galaxy clusters to measure the splashback feature. Studies utilizing X-ray data have shown that the splashback radius can be detected through the density profiles of galaxies surrounding these clusters, thereby offering a complementary perspective to optical and SZ-selected methods \cite{Divya2023, Contigiani2019, Umetsu2017}).

In this paper, we use the cluster sample sourced from the second ROSAT All-Sky-Survey source catalog (2RXS, \cite{Boller16}). We use these X-ray selected galaxy clusters and cross-correlate them with optical galaxies taken from the Dark Energy Survey (DES) Y3 RedMaGiC galaxy catalog \cite{Pandey2021}. We use the DES Y3 shape catalog \cite{Gatti2021} to measure the weak lensing signal around our clusters to obtain an estimate of the halo mass and calculate the traditional spherical over-density boundary $R_{\rm 200m}$.

The splashback radius can be inferred from galaxy number density profiles if galaxies behave as test particles within the cluster's gravitational potential and have a similar dynamic distribution to dark matter particles. Dynamical friction plays a significant role in shaping the splashback radius of galaxy clusters, mainly by influencing the orbits of subhaloes within the cluster's gravitational potential. This effect tends to reduce the measured splashback radius when derived from galaxy number density profiles, as dynamical friction decays the orbits of infalling material, leading to a more compact distribution of galaxies at the cluster outskirts \cite{More2016}. However, \citet{Oshea2024} have shown in simulation that for clusters with masses $M_{\rm 200m} > 10^{14} M_{\odot}$, dynamical friction does not have any significant effect on the measurement of splashback radius as the gravitational binding energy is substantial enough to mitigate the impact of dynamical friction on splashback measurements. In order to assess the role of dynamical friction in shaping the location of the splashback radius, we will follow the methodology in \citet{More2016} and also explore the relationship between the density profiles of dark matter and subhalos hosting galaxies using the MultiDark-Planck II (MDPL2) simulation \cite{mdpl2}. By performing simple subhalo abundance matching, we show that for our fiducial galaxy sample, the location of the steepest slope in the subhalo distribution closely tracks that of the dark matter.

We describe the different data catalogs we use in Section \ref{sec:data}, while the measurement and the modelling techniques are described in Section \ref{sec:model}. In Section \ref{sec:results}, we present the main results and compare them with those of earlier works. We then summarize our findings in Section \ref{sec:conclusion}. Throughout the work, we use a flat $\Lambda$CDM cosmological model with matter density $\Omega_{\rm m} h^2 = 0.27$, baryon density  $\Omega_{\rm b} h^2 = 0.049$, power law index of the initial power spectrum $n_{\rm s}=0.95$, the root mean square variance of density fluctuation $\sigma_8 = 0.81$ and the Hubble parameter $h = 0.7$ as our fiducial cosmological model. The symbol $r$ represents the three-dimensional, while $R$ represents the projected two-dimensional radial distance from the cluster center. We use the halo mass definition of $M_{\rm 200m}$ and corresponding halo boundary $R_{\rm 200m}$ as the radius enclosing a matter density which is 200 times the present matter density of the Universe. The symbol $\log$ will be used to denote logarithm to the base ten.

\section{\label{sec:data}Data}

\subsection{\label{sec:cluster} Galaxy cluster catalog}
ROSAT, the ROentgen SATellite, was the first all-sky X-Ray imaging telescope launched in 1990 \cite{Trumper90}. In 1990-1991, the ROSAT All-Sky Survey (RASS, \cite{Voges99}) was carried out, resulting in significant improvement in the number of known X-ray sources. The cluster sample used in this study is taken from the RASS-MCMF catalog \cite{Klein23}. This catalog is built from the second ROSAT All-Sky-Survey source catalog (2RXS, \cite{Boller16}) by using the Multi-Component Matched Filter cluster confirmation algorithm (MCMF, \cite{Klein23}) on the Legacy Survey DR10 dataset. The RASS-MCMF cluster catalog consists of 8,449 X-ray selected galaxy clusters over a 25,000 $\mathrm{deg}^2$ area of the extra-galactic sky, having a purity of 90\%. More than 90\% of the sky area covered by the survey is not significantly affected by high stellar density or neutral hydrogen density column $N_{\rm H}$.

We convert the RASS-MCMF count rates into flux using the formula $F_{\rm X} = {\rm CR} \times 1.08 \times 10^{-11} \,\text{ergs} \, \text{s}^{-1} \, \text{cm}^{-2}  $, where CR is the count rate. Assuming an underlying power-law spectrum, a correction factor of $1.08$ is used \cite{Boller2016}. Using our cosmological model, we converted the obtained flux to Luminosity $L_{\rm X} = 4\pi d_l^2(z) F_X$, where $d_l(z)$ is the luminosity distance at redshift $z$. We select RASS-MCMF clusters present in the DES RedMaGiC sky area, luminosity cut of $L_{\rm X} > 10^{44} \,\text{ergs} \, \text{s}^{-1} \, \text{h}^{-2}$ and redshift $z < 0.6$. This selection cut results in a sample of 255 X-ray galaxy clusters distributed in an approximate volume-limited manner that we have used in our analysis. We have not used any $k$-correction on our sample.

\begin{figure}
\centering
\includegraphics[width=\columnwidth]{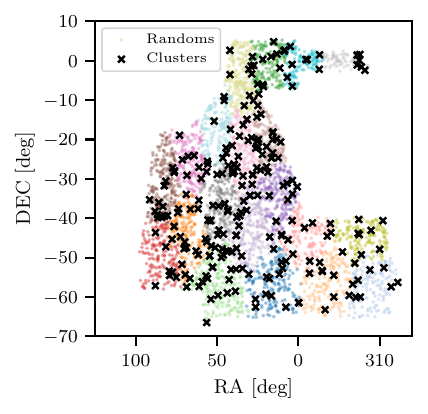}
\caption{The distribution of galaxy clusters and cluster randoms in celestial coordinates. The black crosses represent the galaxy clusters used for our analysis. The colored dots represent the cluster randoms generated, with the colors representing the individual jackknife regions used for covariance estimation.}
\label{fig:sky plot}
\end{figure}

In order to use the Landy-Szalay estimator for the cross-correlation measurements between galaxy clusters and galaxies, we also use galaxy cluster randoms. We randomly distribute points over the DES RedMaGiC survey mask to generate 3888 random galaxy cluster points.
% In order to use the Landy-Szalay estimator for the cross-correlation measurements between galaxy clusters and galaxies, we use galaxy cluster randoms. We randomly distribute points over a 25,000 ${\rm deg}^2$ area of the extra-galactic sky covered by the RASS-MCMF catalog. The number of points distributed over the sky is 20 times the number of galaxy clusters that remain after applying our fiducial luminosity and redshift cuts. The DES RedMaGiC survey mask was applied to this sample to generate 3888 random galaxy cluster points.
Figure \ref{fig:sky plot} shows the on-sky distribution of galaxy clusters and the generated cluster randoms. The different colours of the cluster randoms represent the individual jackknife regions that were used to estimate the errors in our measurements. 

%\begin{figure}[hbtp]

\subsection{\label{sec:galaxy catalog}Galaxy Catalogs}
We use galaxy catalogs from the Dark Energy Survey (DES, \cite{darkenergysurvey}), which is a \~5000 $\mathrm{deg}^2$ optical imaging survey in the Southern Galactic region using the g, r, i, z, and Y bands, with typical magnitude limits of $g \approx 24.3$, $r \approx 24.1$, $i \approx 23.6$, $z \approx 22.8$, and $Y \approx 21.7$. We use the publicly available red-sequence Matched-filter Galaxy Catalog (RedMaGiC) derived from the DES gold galaxy catalog for the calculation of splashback radius using cluster-galaxy cross-correlation and the DES Year 3 shape catalog for weak lensing measurements. 
\subsubsection{\label{sec:RedMaGiC}DES Y3 RedMaGiC galaxy catalog.}
To calculate the splashback radius, we use galaxies selected by the RedMaGiC algorithm \cite{Rozo2016} on the DES Y3 data.  The RedMaGiC algorithm fits each galaxy to a template corresponding to red sequence galaxies in order to determine the best-fitting redshift $z_{\rm photo}$, which is then used to compute the galaxy luminosity $L$. The RedMaGiC catalog \cite{Pandey2021} is then constructed by selecting bright galaxies ($L\ge L_{\rm min}$) and galaxies that had a good fit to the red-sequence template ($\chi^2 / {\rm dof} \le 2$). Of the two RedMaGiC catalogs that were constructed in this manner, we use the ``high density" catalog which is a volume-limited sample to a $z$-band luminosity of $0.5\, L_*$, and this catalog is divided into three tomographic bins of $z \in (0.15, 0.3], (0.3,0.5], (0.5,0.65]$ and has a limiting $z$-band magnitude limit of $m_z=21.3$. The DES Y3 RedMaGiC random catalog is generated by uniformly distributing points over the survey footprint defined by the mask and assigning each point a redshift drawn from the tomographic distributions. The random catalog is 40 times larger than the galaxy catalog.

\subsubsection{\label{sec:desy3} DES Y3 Shape catalog }
The DES Y3 shape catalog encompasses over 100 million galaxies observed using the DECam instrument in Chile. It spans approximately $5000$ sq deg of the southern sky, capturing photometric data in the $griz$ bands. Galaxy shapes in the DES Y3 catalog are obtained using the \textsc{metacalibration} algorithm presented in \citet{Huff2017} and \citet{Sheldon2017}. 

\textsc{metacalibration} measures the shapes of galaxies in the $riz$ band of DES Y3 data. The DES Y3 shear catalogue has a source number density of $5.59 \,\text{arcmin}^{-2}$ and shape noise of $0.261$. The shape of every galaxy is defined as a two-component ellipticity, $e$ as detailed in \citet{Gatti2021}. The average of measured shapes of galaxies can be used to estimate the true shear imparted on them due to gravitational lensing, $\gamma$, by accounting for the response matrix, $\mathcal{R}$, such that
\begin{equation}
\langle \gamma \rangle = \langle \mathcal{R} \rangle ^{-1} \langle e \rangle\,.
\end{equation}
Here, the angled brackets denote an ensemble average, and $\mathcal{R}$ represents the response of the measured galaxy shapes to the true shear. \textsc{metacalibration} calculates the additional selection response term $\mathcal{R}_{\text{s}}$ as the galaxy shapes depend on the specific sample selection. The quantity $\mathcal{R}_{\text{s}}$ represents the response of measured shapes to an input shear specific to the selected galaxy sample. Thus the total response becomes $\mathcal{R}+\mathcal{R}_{\text{s}}$ which is a $2 \times 2$ matrix, but it is well represented by the average of the diagonal components (following, e.g., \cite{Prat2018}). 

Since we are interested in the shapes of the cluster galaxy density profile, we will ignore any per cent level multiplicative shear biases, such as those arising from blended galaxies \cite{Sheldon2020, MacCrann2021}, inferred from image simulations. Any such corrections are subdominant in our analysis, as the errors on the measurement of the splashback radius itself dominate our error budget.
% \textsc{metacalibration} algorithm does not fully take into account blended galaxies, where one single detection is associated with multiple, unresolved galaxies \cite{Sheldon2020, MacCrann2021}. Since we are interested in the shapes of the cluster density profile, we will ignore such per cent level multiplicative shear biases inferred from image simulations. Any such corrections are subdominant in our analysis, as the errors on the measurement of the splashback radius itself dominate the error budget.

\section{\label{sec:model}Modelling and Measurements}

\subsection{\label{sec:wl_model}Weak Lensing Profile}
Gravitational lensing refers to the distortion of light from distant background galaxies caused by the gravitational influence of intervening correlated structure in the matter distribution. Weak lensing produces small coherent distortions in the shapes of galaxies that can only be detected statistically (see \cite{Kilbinger2015, Mandelbaum2018, dodelson2017book}), for a recent review). Weak lensing analysis typically concerns itself with the shear, which is expressed as a complex quantity that describes the distortion of the original shape of a background galaxy. The tangential shear of a background galaxy, which is the distortion in the direction tangential to the line joining the lens and source galaxy is given by;
\begin{equation}
\gamma_t = -\gamma_1 \text{cos}2\phi - \gamma_2 \text{sin} 2 \phi,
\end{equation}
where $\phi$ is the position angle of the source galaxy with respect to the $x$-axis of the system, $\gamma_1$ and $\gamma_2$ are the shear components in a cartesian coordinate system. 

This tangential shear is related to the projected matter distribution of the halo given by
\begin{equation}
\gamma_t = \dfrac{\overline{\Sigma}(R) - \langle\Sigma(R) \rangle}{\Sigma_{\rm crit}}\,,
\end{equation}
where, the numerator, $\overline{\Sigma}(R) - \langle{\Sigma(R)} \rangle = \Delta\Sigma(R)$ is called the excess surface mass density (ESD). The quantity $\bar{\Sigma}(R) = \frac{\int_{0}^{R} \Sigma(R') 2\pi R' \, dR'}{\pi R^2}$ is the average surface matter density within a projected distance $R$, while $\langle{\Sigma(R)} \rangle$ is the azimuthally averaged surface matter density at the same distance. The critical surface mass density $\Sigma_{\rm crit}$ is given by;
\begin{equation}
\Sigma_{\text{crit}} = \frac{\frac{c^2}{4\pi G} D_a(z_s)}{(1 + z_l)^2 D_a(z_l) D_a(z_l, z_s)}
\end{equation}
Here, $D_a(z_s)$, $D_a(z_l)$ and $D_a(z_l, z_s)$ are the angular diameter distances between the observer and the source at redshift $z_s$, the observer and lens at redshift $z_l$ and between the lens-source pair, respectively. The factor $(1 + z_l)^2$ arises due to our use of comoving coordinates. 

We can measure the $\Delta\Sigma(R)$ for our lens sample using the estimator
\begin{equation}
\Delta\Sigma(R) = \frac{\sum_{ij}s^{ij}e_{\mathrm{t}}^{ij}(R)}{\sum_{ij}s^{ij}\Sigma_{\mathrm{c,MC}}^{-1}\left(z_{1}^{i},z_{\mathrm{s}}^{j}\right)(\mathcal{R}^{j}+\mathcal{R}_{\mathrm{s}})}\,,
\end{equation}
where $i$ and $j$ represent the lenses and sources, respectively. $\mathcal{R}$ is the shear response from \textsc{metacalibration}; $\mathcal{R}_s$ is the selection response, and $s^{ij}$ is the weight applied given by;
\begin{equation}
s^{ij}=\omega^j\Sigma_{\mathrm{c,mean}}^{-1}
\begin{pmatrix}
z_1^i,z_\mathrm{s}^j\,.
\end{pmatrix}
\end{equation}
Here $\omega^j$ is the square inverse of the measured shear uncertainty of the source (see \cite{Gatti2021}) for detailed information). Also, $\Sigma_{\mathrm{c,MC}}^{-1}=\Sigma_{\text{crit}}(z_l, z_s^{\text{MC}})^{-1}$ is the inverse critical density with the source redshift $z_s^{\text{MC}}$ randomly chosen from the photo-$z$ distribution estimated with Bayesian Photometric Redshift (BPZ), and $\Sigma_{\mathrm{c,mean}}^{-1}$ is the one evaluated at the mean redshift from BPZ. See \citet{McClintock2018} for a detailed description of the estimator. The BPZ redshift uncertainties could introduce upto percent level differences in the lensing signals (see e.g., \cite{McClintock2018}), which are subdominant given the current statistical uncertainties in the measurements of the splashback radius.

We used the stacked weak lensing signal of all clusters and calculate $\Delta\Sigma(R)$ in ten projected logarithmically-spaced bins from the cluster center with a comoving distance in the range of $R\equiv [0.2, 12] h^{-1}\text{Mpc}$. We exclude the region below $0.2 h^{-1} \text{Mpc}$ as the crowding of galaxies near the cluster center makes robust calculation of background shear difficult so that $\Delta\Sigma(R)$ becomes uncertain. The cross component of the signal $\Delta\Sigma_{\times} = \gamma_{\times} \Sigma_c (z_l, z_s)$ is zero by parity and thus is a useful measure of presence of systematics in the weak-lensing measurements \cite{Schneider2005}. We thus use the measurement of this cross-component signal as a null test.

We also calculate $\Delta\Sigma(R)$ in the same ten logarithmic bins around our stacked cluster randoms. The weak lensing signal around random points is expected to be around zero as there would be no significant mass concentrations along the line of sight. Since we do not find substantial evidence for a non-zero signal around random points, we do not subtract this signal from our measurements. 

We applied a redshift-dependent selection criterion of $z_s > z_l+\Delta z$ for the source galaxies to ensure a cleaner population of background galaxies with $\Delta z = 0.3$, and $z_s$ and $z_l$ are the redshift of the source and the lens galaxies, respectively. This approach helps us in controlling for source galaxies that may be physically associated with our lens sample, which could dilute the weak lensing signal measurements. Additionally, we account for any remaining associated galaxies by using the boost factor \( C(R) \) \cite{SDSS:2003slg, Mandelbaum:2008iz}. This factor is calculated by taking the ratio of weighted lens-source pairs to weighted random-source pairs within a given radial bin \( R_i \) and is given by
\begin{equation}
\label{eq:boost}
    C(R_i) = \frac{N_{\rm r}}{N_{\rm l}}\frac{\sum_{ij} s^{\rm ij}}{\sum_{\rm kj} s^{\rm kj}},
\end{equation}
where the summations over \(i\), \(j\), and \(k\) correspond to lenses, sources, and randoms, respectively, the terms \(N_{\rm l}\) and \(N_{\rm r}\) represent the number of lenses and random, respectively.

We measured \(C(R)\) in radial bins using randoms that are 20 times larger than our lens sample and we calculated the associated errors using 20 area jackknife realizations. These randoms are constructed to match the same sky coverage, adhere to the same star mask, and maintain a similar redshift distribution as our lens galaxies. In Appendix \ref{App:boost parameter}, we demonstrate that our measurements of the boost parameter are consistent with unity, given the statistical uncertainties. Therefore, we neglect the boost parameter in our analysis.

Our boundary definition, $r_{\rm 200m}$ is the three-dimensional radius of the cluster, which encloses a density of 200 times the present matter density of the Universe, and hence is given by, 
\begin{equation}
r_{200\text{m}} = \left( \frac{3M_{200\text{m}}}{4\pi  200 \rho_{\rm m}} \right)^{1/3}.
\end{equation}

In order to model the weak lensing signal we start from the Navarro-Frenk-White (NFW) density profile \cite[][]{NFW_profile}, $\rho_{\text{nfw}}(r)$,
\begin{equation}
\rho_{\text{nfw}}(r) = \frac{\delta_c \rho_{\rm m}}{\left( \frac{r}{r_{\rm s}} \right) \left( 1 + \frac{r}{r_{\rm s}} \right)^2},
\end{equation}
such that
\begin{equation}
\delta_c = \frac{200}{3} \frac{c^3}{\ln(1+c) - c/(1+c)}\,,
\end{equation}
and $r_{\rm s} = r_{\rm 200m}/c$ is the scale radius of the halo. 

With just two parameters $\Theta = (M_{\rm 200m},c)$, the halo mass $M_{\rm 200m}$ and concentration parameter $c$, we can predict the ESD profile $\Delta\Sigma(R)$ for a set of our model parameters. In practice, we use the analytical form for $\Delta\Sigma(R)$ for the case of the NFW profile as given by eqn. 14 in \citet{wright1999}
\begin{equation}
\Delta\Sigma(R) = r_{\rm s} \delta_c \rho_{\rm m} f(R / r_{\rm s}),
\end{equation}
\begin{equation}
f(x) =
\begin{cases} 
g_{<}(x), & x < 1, \\
\frac{10}{3} + 4 \ln\left(\frac{1}{2}\right), & x = 1, \\
g_{>}(x), & x > 1,
\end{cases} \tag{13}
\end{equation}
where the functions $g$ are given by
\begin{equation}
\begin{split}
g_{<}(x) &= \frac{8 \, \text{arctanh} \sqrt{(1-x)/(1+x)}}{x^2 \sqrt{1-x^2}} 
+ \frac{4}{x^2} \ln\left(\frac{x}{2}\right) \\
&\quad - \frac{2}{x^2 - 1} 
+ \frac{4 \, \text{arctanh} \sqrt{(1-x)/(1+x)}}{(x^2-1)(1-x^2)^{1/2}},
\end{split}
\end{equation}
and 
\begin{equation}
\begin{split}
g_{>}(x) = \frac{8 \, \text{arctanh} \sqrt{(x-1)/(1+x)}}{x^2 \sqrt{x^2-1}} 
+ \frac{4}{x^2} \ln\left(\frac{x}{2}\right) \\
- \frac{2}{x^2 - 1} 
- \frac{4 \, \text{arctanh} \sqrt{(x-1)/(1+x)}}{(x^2-1)^{3/2}}.
\end{split}
\end{equation}
As seen in Table \ref{tab:model_para}, during our inference of the posterior distribution of parameters given the weak lensing measurements, we adopt flat priors on our parameters. 

The dark matter contribution is the dominant component, so we have not added a point mass contribution for the baryonic component of the central galaxy of the cluster as done by some of the previous studies (e.g., \cite{Kobayashi2015}). We also do not split the dark matter contribution into 1-halo and 2-halo terms (for more details, see \cite{Hikage2013, Miyatake2016}) or use an off-centering kernel \cite{Johnston2007} in order to account for the possible misidentification of the cluster center. None of these effects are expected to cause a significant difference to our conclusions, given the scales we fit and the statistical errors on our mass estimate. We infer the average halo mass rather than the entire distribution of the halo masses, and hence we do not carry out a halo occupation distribution (HOD) based modelling (e.g., \cite{Seljak2000, Cooray2002, Bosch2013}). The distribution of our galaxy cluster with such high masses is expected to peak due to the presence of the exponential tail of the mass function. We thus limit our modelling to the simple NFW profile described above. Using numerical simulations of our selection of galaxy clusters, we will show that this approach yields a sufficiently accurate prediction for the location of the splashback radius given the accuracy of our measurement.

Along with the NFW profile, we also fit our weak lensing measurements $\Delta \Sigma (R)$ using the profile suggested by \citet{Diemer_Kravtsov_2014} (DK14 profile) as explained in section \ref{sec:dk14}. This profile also includes a two halo term in addition to the one halo term. Although our weak lensing measurements alone were not able to constrain the location of the splashback radius, we found that the mass estimate obtained using the DK14 profile are consistent with those obtained from the NFW profile. In future, we expect to calculate the splashback radius of X-ray clusters using the weak lensing measurements (like the SZ cluster study by \cite{Shin2021}).

\subsection{\label{sec:dk14} Galaxy Number Density Profile}
For our cluster galaxy cross-correlation measurements, we use the method developed by \citet{More2016}. The galaxy catalogs we use provide red sequence based photometric redshifts, and we avoid using them to determine the absolute magnitude of galaxies. We assign magnitudes to galaxies based on the redshift of the cluster against which we cross-correlate. In our fiducial analysis, we include the galaxy in the pair counts if the galaxy has a $z$-band magnitude $M_{z}-5 \log h$ brighter than $-21.2$. This allows us to statistically measure the cross-correlation between galaxy clusters and galaxies brighter than a given magnitude limit in a consistent manner for all of our sample. We use the Landy-Szalay estimator \cite{1993Landy}, which is less prone to systematics as compared to other estimators, especially for large scales \cite{Kerscher_2000}. The projected cross-correlation $\xi_{\rm 2D}(R)$ at a projected comoving radius is given by
\begin{equation}
    \label{eq:lz}
    \xi_{\rm 2D}(R) = \frac{ D_{\rm c} D_{\rm g} - D_{\rm c} R_{\rm c} - R_{\rm c} D_{\rm g} + R_{\rm c} R_{\rm g}}{ R_{\rm c} R_{\rm g} }
\end{equation}
Where $D_{\rm c} D_{\rm g} , \, D_{\rm c} R_{\rm c} ,\, R_{\rm c} D_{\rm g} ,$ and $ R_{\rm c} R_{\rm g}$ are the pair counts between clusters-galaxies, clusters-galaxy randoms, cluster randoms-galaxies, and cluster randoms-galaxy randoms, respectively. The difference between the number of random points and the corresponding number of galaxies or clusters are normalized appropriately while calculating the cross-correlation $\xi_{\rm 2D}(R)$. We measure the signal over nine logarithmically spaced comoving projected radial bins from 0.2 to 8.0 $\text{h}^{-1} \text{Mpc}$. 

We use the DK14 profile in order to model the cross-correlation signal of the density profile, such that
\begin{align}
%\label{eq:3drsp}
\label{eq:xi3d}
\xi_{\rm 3D}(r) &= \rho_{\text{in}}(r) f_{\text{trans}}(r) + \rho_{\text{out}}(r),  \\
\label{eq:rhoin}
\rho_{\text{in}}(r) &= \rho_{\rm s} \exp \left( -\frac{2}{\alpha} \left[ \left( \frac{r}{r_{\rm s}} \right)^\alpha - 1 \right] \right),  \\
\label{eq:rhout}
\rho_{\text{out}}(r) &= \rho_0 \left( \frac{r}{r_{\text{out}}} \right)^{-s_e}, \\
\label{eq:ftrans}
f_{\text{trans}}(r) &= \left( 1 + \left( \frac{r}{r_{\rm t}} \right)^\beta \right)^{-\frac{\gamma}{\beta}}.  
\end{align}
Here, the density profile $\xi_{\rm 3D}$ consists of an inner Einasto profile \cite{Einasto1965} $\rho_{\rm in}$ multiplied by a transition function $f_{\rm trans} $ at the location of the splashback radius and the outer profile $\rho_{\rm out}$ given by a power law. 

We compute the two-dimensional cross-correlation profile \(\xi_{\rm 2D}(R)\) at the projected radius \(R\) by integrating the three-dimensional profile \(\xi_{\rm 3D}(r)\) along the line of sight direction \(\pi\),
\begin{equation}
\label{eq:2dcorr}
\xi_{\rm 2D}(R) = \frac{1}{R_{\text{max}}} \int_0^{R_{\text{max}}} \xi_{\rm 3D} \left( \sqrt{R^2 + \pi^2} \right) d\pi. %\tag{17}
\end{equation}
Equation \ref{eq:2dcorr} comprises nine parameters - $\rho_{\rm s}, \alpha, r_{\rm s}, \rho_0, r_{\text{out}}, s_{\rm e}, \beta, r_{\rm t}, \gamma$. As $\rho_0$ and $r_{\text{out}}$ are degenerate, we fix the $r_{\text{out}} = 1.5 \, \text{h}^{-1} \, \text{Mpc}$. Thus, we use the eight free parameters $\Theta = (\rho_{\rm s}, \alpha, r_{\rm s}, \rho_0, s_{\rm e}, \beta, r_{\rm t}, \gamma)$ to model the data. We fix the maximum projected length $R_{\text{max}} = 40 \, \text{h}^{-1} \, \text{Mpc}$ following \citet{More2016}. We do not use photometric redshifts for galaxies, as the cross-correlation with cluster centers naturally selects the physically associated galaxy population.

The priors on the eight parameters used for inferring the posterior distribution of parameters given our measurements are listed in Table \ref{tab:model_para}.
\begin{table}[th]
\centering
\begin{tabular}{ll}
\toprule
\multicolumn{2}{c}{\textbf{Model Parameters: DK14}} \\
\midrule
\textbf{Parameter} & \textbf{Prior} \\
\midrule
$\log \rho_{s, \mathrm{GP}}$ & Flat[-3, 5] \\
$\log \rho_{s,\mathrm{WL}}$ & Flat[-5, 5] \\
$\log \alpha$ & Gauss(log(0.2), 0.6) \\
$\log r_{\rm s}$ & Flat[log(0.1), log(5.0)] \\
$\log \rho_0$ & Flat[-1.5, 1.5] \\
$s_{\rm e}$ & Flat[0.1, 4] \\
$\log r_{\rm t}$ & Flat[log(0.5), log(1.6)] \\
$\log \beta$ & Gauss(log(6.0), 0.2) \\
$\log \gamma$ & Gauss(log(4.0), 0.2) \\
\\
\midrule
\multicolumn{2}{c}{\textbf{Model Parameters: NFW}} \\
\midrule
\textbf{Parameter} & \textbf{Prior} \\
\midrule
$\log(M_{\rm 200m}/h^{-1} M_{\odot})$ & Flat[12,16] \\
$c$ & Flat[0,20] \\
\bottomrule
\end{tabular}
\caption{\textit{Parameter priors:} The table above gives the prior distribution used for the MCMC chains on our model parameters. The Flat[a,b] represents uniform priors in the interval (a,b), and Gauss($\mu$,$\sigma$) denotes the Gaussian priors with $\mu$ mean and $\sigma$ standard deviation. The top part of the table shows the prior distribution used for the DK14 model parameters. The subscripts GP and WL in the parameter $\log \rho_{s}$ represent the prior distribution used for galaxy number density and Weak lensing data, respectively. The halo mass $\log [M_{\mathrm{200m}}/h^{-1} \mathrm{M}_\odot] $ and concentration $c$ are for the NFW profile-based modelling of the weak lensing measurements for halo mass calibrations.} 
\label{tab:model_para}
\end{table}

\subsection{\label{sec:covariance}Covariance estimation}
For accurate and unbiased covariance estimation, we need to consider the spatial variations in our measurements. Since our datasets exhibit strong pairwise agreement between measurements (i.e., the agreement between different observational regions), the jackknife technique \cite{Miller_1974} offers a robust and computationally efficient method for variance estimation. In this technique, we first divide our clusters and cluster randoms into 20 distinct jackknife regions as shown in Figure \ref{fig:sky plot}. We iteratively exclude clusters from one jackknife region at a time and recalculate the measurements for the resampled datasets. The covariance of our measurement, obtained via the jackknife resampling technique, is calculated as
\begin{equation}
    C_{ij} = \frac{N - 1}{N} \sum_{k=1}^{N} (\mathcal{X}_k^i - \langle\mathcal{X}^i\rangle)(\mathcal{X}_k^j - \langle\mathcal{X}^j\rangle)
\end{equation}
where $i,j$ are the radial bins, $N$ is the number of jackknife regions, and $\mathcal{X}_k$ is our measurement of $k$th jackknife resampled data. $ \langle\mathcal{X}^i\rangle$ represents the mean value across all the jackknife samples. In case of our weak lensing analysis, our measurement $\mathcal{X}$ would be the excess surface mass density $\Delta\Sigma$, while for our cluster-galaxy analysis, $\mathcal{X}$ would be the 2D cross-correlation measurement $\xi_{\mathrm{2D}}$.  We use eqn. 17 in \citet{Hartlap2007} in order to account for the bias due to the noise in the covariance estimation. The errors on our measurement are calculated using this jackknife resampled covariance values. Finally, we obtain the correlation matrix by normalizing each element of our covariance matrix $C_{ij}$ with the diagonal uncertainties
\begin{equation}
    r_{\rm ij} = \frac{C_{\rm ij}}{\sqrt{C_{\rm ii}C_{\rm jj}}}
\end{equation}
where subscripts ij represent the $\text{i}^{\rm th}$ and $\text{j}^{\rm th}$ radial bins. The correlation matrix for our weak lensing and galaxy number density measurements is shown in Appendix \ref{App:covariance}.

\subsection{\label{sec:mcmc} Model Fitting}
Given the data $\mathcal{D}$, we carry out a Bayesian analysis in order to estimate the posterior probabilities for our model parameters $\Theta$. As seen in table \ref{tab:model_para}, we use Gaussian priors on $\log \alpha$, $\log \beta$, and $\log \gamma$ with flat priors on rest of the parameters. In our work, we perform separate fits for weak lensing and projected galaxy number density profiles. The priors used for model fitting are similar to those mentioned in the literature; \cite{More2016, Murata2020, Divya2023} for galaxy number density profile and \cite{Chang2018, Shin2019, Shin2021} for weak lensing profile and are motivated from results from numerical simulations \cite{Diemer_Kravtsov_2014, Gao2008}.

From the Bayes theorem, we can write the posterior distribution of parameters as
\begin{equation}
P(\Theta \mid \mathcal{D}) \propto P(\mathcal{D} \mid \Theta) P(\Theta) ,
\end{equation}
where $P(\mathcal{D} \mid \Theta)$ is the likelihood of the data given model parameters. We use Gaussian likelihood given by
\begin{equation}
P(\mathcal{D} \mid \Theta) \propto \exp\left(-\frac{\chi^2(\Theta)}{2}\right)
\end{equation}
where $\chi^2(\Theta) = \left[ \mathcal{D} - \mathcal{M}(\Theta) \right]^T C^{-1} \left[ \mathcal{D} - \mathcal{M}(\Theta) \right]$, $\mathcal{D}$ is the data vector with $\mathcal{M}(\Theta)$ as the model prediction vector given the parameter $\Theta$. We use the affine invariant Monte Carlo Markov Chain (MCMC) sampler of \citet{Goodman2010} as provided by the python package \textsc{emcee} \cite{Foreman-Mackey_2013} to obtain the posterior distribution of model parameters. In our work, we run MCMC to obtain samples from the posterior distribution of parameters given our measurements of the weak lensing signal. We also run MCMC to obtain samples from the parameter posteriors for our inference of the splashback radius in the cross-correlation measurements of our cluster sample. For DK14 profile modelling, we used 512 walkers with 3000 steps, and the chains converged within 1000 steps. For the NFW fit to the weak lensing profile, we use 64 walkers with 45000 steps where the chains converged within 5000 steps. The posterior distribution of the halo mass for the weak lensing fits and the 3-D splashback radius for the cross-correlation fits did not show any significant variations after the chains have converged.

\section{\label{sec:results}Results}

\subsection{\label{sec:weak_lensing} Halo mass for galaxy clusters}

\begin{figure}[hbtp]
\centering
\includegraphics[width=\linewidth]{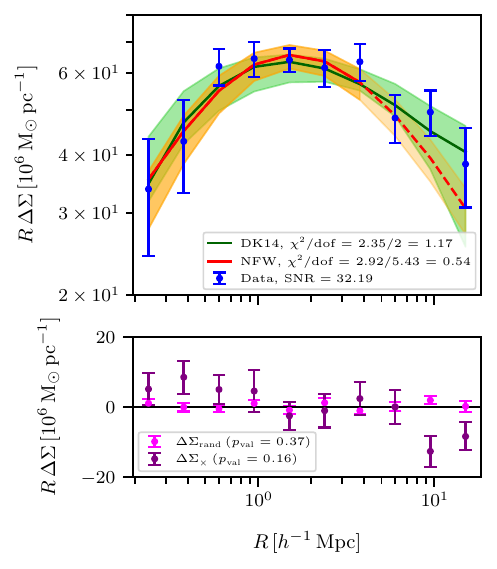}
\caption{\textit{Weak Lensing signal}: Blue points with errors in the top panel correspond to the measured weak lensing signal and the shape noise errors for our X-ray selected cluster sample. The orange solid curve shows the best fit NFW model prediction to our data with $\chi^2_{\rm red}\equiv\chi^2/\text{dof}$ listed in the legend. The dashed orange line represents the extrapolated NFW model predictions without the two-halo term.  The shaded orange region represents the 68 percentile around the median model predictions. The solid green line corresponds to the DK14 model fit to the data, with the shaded region showing the 68 percentile around this median model prediction. The cross-component of the weak lensing signal, $\Delta\Sigma_{\times}$, with shape noise errors, and the signal around random points $\Delta\Sigma_{rand}$, with errors over mean computed using the random realisations are shown in the bottom panel.} %The best fit NFW model parameter values are listed in Table \ref{tab:model_constraints}. These are both consistent with zero.}
\label{fig:wl_fit}
\end{figure}

\begin{figure}[hbtp]
\centering
\includegraphics[width=\linewidth]{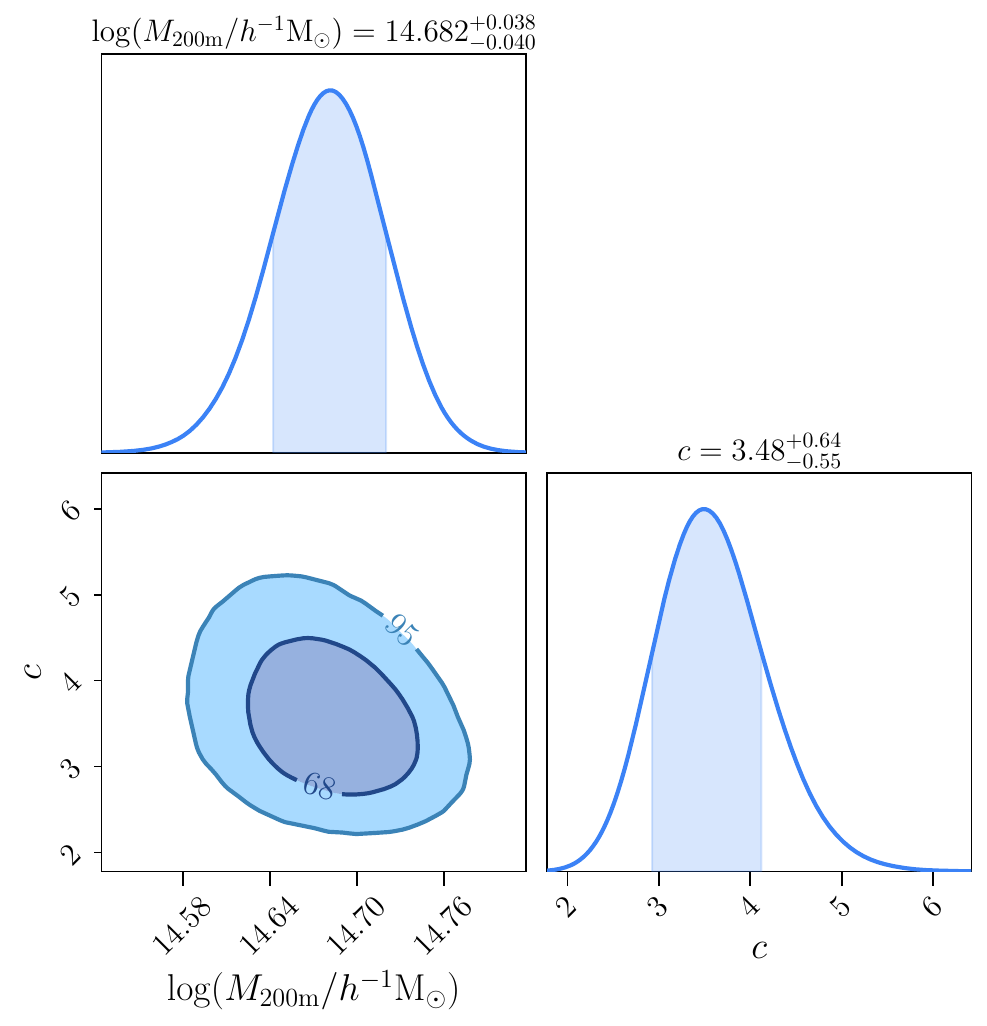}
\caption{\textit{Posterior of the mass profile parameters fiven the weak lensing signal}: The triangle plot shows the degeneracy between the cluster halo mass $\log M_{\rm 200m}$ and concentration parameter $c$, respectively. The light and dark blue shaded contours are 68 and 95 credible regions.}
\label{fig:wl_contours}
\end{figure}

In this section, we present the result and analysis of weak lensing analysis around our X-ray selected galaxy clusters. We use the methodology described in section \ref{sec:wl_model}; we measure the weak lensing around our galaxy cluster in ten projected logarithmically-spaced bins from the cluster centre with comoving distance in the range of [0.2, 12] $h^{-1} \text{Mpc}$. We infer the halo mass $M_{\rm 200m}$ and the corresponding spherical overdensity size $r_{\rm 200m}$ by modelling the data with the NFW profile. We also model our signal using the DK14 profile to check the consistency of the measured halo mass with the NFW profile.

In the top panel of Figure \ref{fig:wl_fit}, blue points represent the measured weak lensing signal $\Delta \Sigma$. The error bars were calculated using the jackknife technique with the data divided into 20 regions as shown in Figure \ref{fig:sky plot} with the signal-to-noise ratio of $32.19$. The bottom panel of Figure \ref{fig:wl_fit} shows the results of the systematic tests. The magenta color points represent the signal for random points $\Delta\Sigma_{\rm rand}$. We use the random points generated as mentioned in Section \ref{sec:cluster}. The p-value obtained for these random points is 0.37. The purple points show the cross-component $\Delta\Sigma_{\times}$ of the signal. The p-value for these points is 0.16. Both these sets of results point towards a null signal, which shows that our weak lensing measurements are robust. 

The solid orange curve shown in the top panel correspond to the best fit NFW profile to the data and has a reduced chi-squared of $\chi^2_{\rm red} = 2.95/5.43$ with the effective degrees of freedom ${\rm dof_{eff}}=5.43$ (see eqn. 29 of \cite{Raveri2019}). Note that we only use the first seven data points shown in the figure to calculate the NFW fit posteriors, as the NFW profile primarily describes the inner regions of halos and does not capture the two halo term at large distances. The dashed orange line represents the extrapolated NFW profile from the best-fit model without the two-halo term. The shaded orange region represents the 68 percentile of the model predictions.

The result of the NFW model fitting to data is shown in figure \ref{fig:wl_contours}. The light and dark regions show the 95 and 68 per cent credible region for the halo mass and concentration parameter of our galaxy clusters. Our weak lensing measurements allow an inference of the halo mass of $\text{log} (M_{\rm 200m}/h^{-1}\text{M}_{\odot}) = 14.682^{+0.038}_{-0.040}$ and a concentration parameter of $c = 3.48^{+0.64}_{-0.55}$. These are statistical uncertainties and we have ignored the systematic uncertainties stemming from imperfections of the shape catalog as their impact is not expected to impact our conclusions given the large error on the position of the splashback radius.

In the top panel of Figure \ref{fig:wl_fit}, the green color line represents the best fit DK14 profile to our weak lensing data. We obtain a reduced chi-squared of $\chi^2_{\rm red} = 2.35/2$ with two degrees-of-freedom. The shaded green color region represents the credible intervals of the measurements. Using equation \ref{eq:xi3d} of the DK14 profile, we calculate the halo mass at which the density of cluster becomes 200 times the critical density of the universe, $\text{log} (M_{\rm 200m}/h^{-1}\text{Mpc}) = 14.658^{+0.026}_{-0.026}$. This value is consistent with the value obtained from the NFW fit to the data. At larger radii, the NFW and DK14 fit diverge as the NFW extrapolated values do not include the 2-halo term.

Based on this inferred halo mass, we derived the corresponding value of the traditional halo boundary $r_{\rm 200m}=1.97^{+0.05}_{-0.06} h^{-1} \text{Mpc}$. We use this constraint on $r_{\rm 200m}$ and compare it with the value of the splashback radius in section \ref{sec:splashback_result}. Any issues related to miscentering are not expected to cause a large change in the expected halo mass (see comparison between results in \cite{More2016, Baxter2017}).

\subsection{\label{sec:splashback_result} Galaxy number density profile}

In this section, we describe our inference of the splashback radius $r_{\rm sp}$ of our galaxy clusters. We use the methodology described in Section \ref{sec:dk14} to measure the cross-correlation signal between our RASS-MCMF galaxy clusters and DES RedMaGic galaxies. This cross-correlation signal is calculated in nine logarithmically spaced comoving projected radial bins in the range [0.2,8.0] $h^{-1}$ Mpc. Using equation \ref{eq:2dcorr}, we model the projected cross-correlation profile $\xi_{\rm 2D}(R)$ and determine the radial position of the splashback feature $R_{\rm sp}$. Furthermore, we infer the corresponding three-dimensional cross-correlation profile $\xi_{\rm 3D}(r)$ and estimate the value of the splashback radius $r_{\rm sp}$.

\begin{figure*}[p]
\centering
\includegraphics[width=\textwidth]{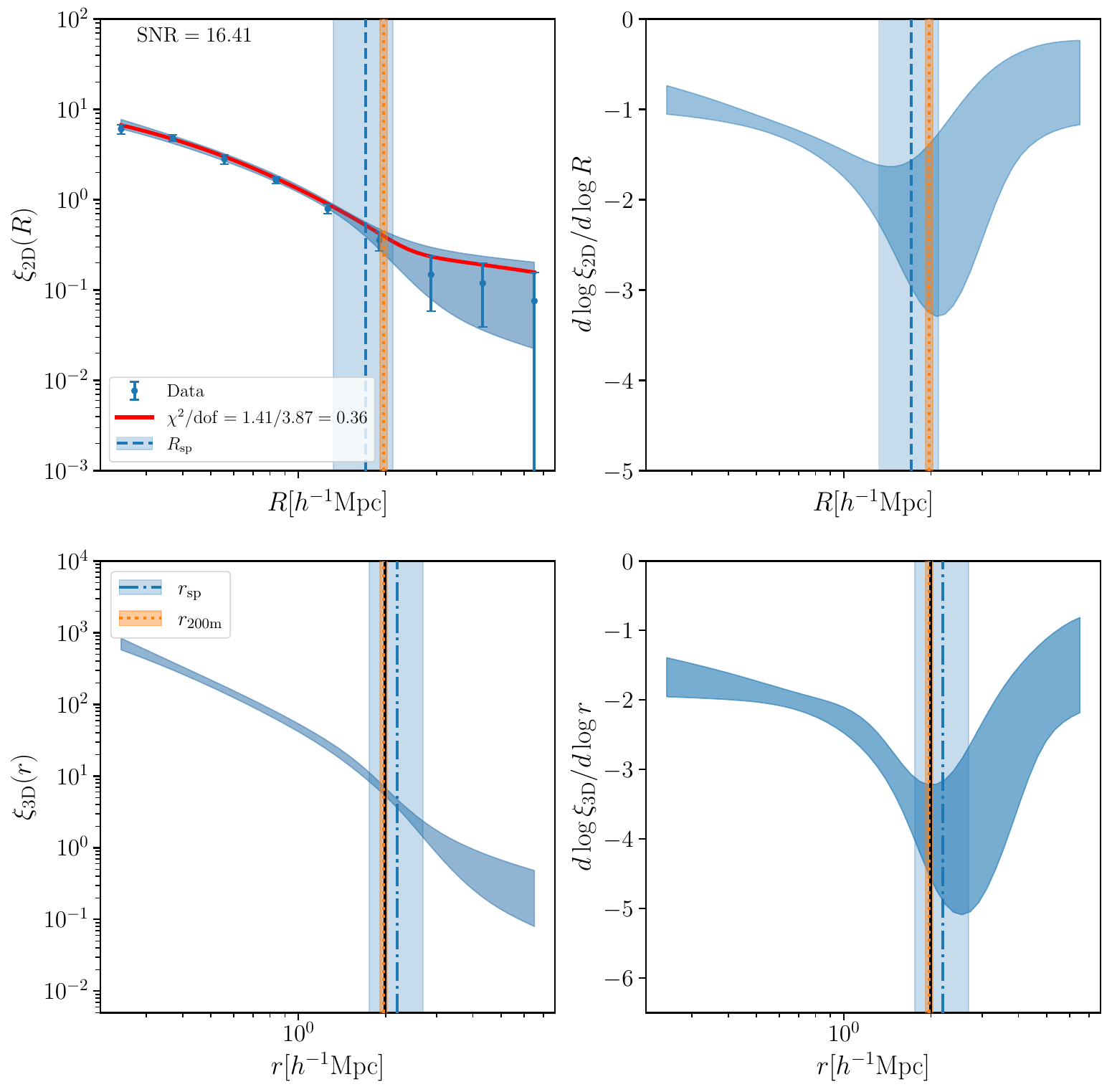}
\captionsetup{justification=justified, width=\textwidth}
\caption{\textit{Galaxy number density profile:} The blue data points in the top left panel shows the measurements of the 2D cross-correlation $\xi_{\rm 2D}(R)$ between the RASS-MCMF X-ray galaxy clusters and the DES Y3 optical galaxies. The solid red line represents our best-fit model predictions. The dark blue shaded regions denote the 68th percentile around the median model predictions. The top right panel shows the inferred logarithmic derivative of the model predictions for the projected cross-correlation profile $\xi_{\rm 2D}(R)$, with the blue dashed vertical line indicating the median value of the inferred two-dimensional splashback radius $R_{\rm sp}$, and a light blue shaded 68th percentile error region around it. The bottom panel shows the corresponding three-dimensional $\xi_{\rm 3D}(r)$ profile and the associated logarithmic derivative. In the bottom row, the blue dot-dashed vertical line and light blue shaded region indicate our constraints on the three-dimensional splashback radius $r_{\rm sp}$. In all panels, the orange dotted vertical line with a shaded region represents the weak lensing-calibrated three-dimensional spherical overdensity size $r_{\rm 200m}$, derived from source galaxies in the DES Y3 shape catalog. The solid black vertical line shows the expected value for $r_{\rm sp}$ based on $\Lambda$CDM predictions \cite{More2015}. Table \ref{tab:model_constraints} provides the model parameter constraints, including the splashback radius and spherical overdensity sizes.}
\label{fig:splashback}
\end{figure*}

The data points with error bars in the top left graph of Figure \ref{fig:splashback} are the cross-correlation measurement values. The cross-correlation signal $\xi_{\rm 2D}(R)$ is measured with a signal-to-noise ratio of $16.41$, where the errors are calculated using the jackknife technique. The red solid line is the best-fit model to the data having a reduced chi-square value of $\chi_{\rm red}^{2} = 0.36$ with the effective degrees of freedom ${\rm dof_{eff}}= 3.87$ (see eqn. 29 of \cite{Raveri2019}). The blue-shaded region represents the 68 per cent credible region. The measurements are in reasonable agreement with the expectations from the model. In the top two graphs of figure \ref{fig:splashback}, the blue dashed vertical line represents the calculated projected splashback radius $R_{\rm sp}$, and the blue shaded region is the 68 percentile error region. The orange vertical dashed line represents the boundary of the halo in the traditional sense - $r_{\rm 200m}$ calculated from our weak lensing analysis as described in section \ref{sec:weak_lensing}.

The top right panel in figure \ref{fig:splashback} shows the credible prediction for the logarithmic slope $d \log(\xi_{\rm 2D}) /d \log(R)$ from the posterior of our model parameters and the blue shaded region represents the 68 per cent credible region to the median. The quantity $R_{\rm sp}^{\rm 2D}$ corresponds to the location of the steepest slope of the projected correlation function. Table \ref{tab:model_constraints} shows the posteriors of the model parameters and the splashback radius along with the 68 percentile error estimates. We obtained a 20 percent constraints on $R_{\rm sp}$ with the median value of $1.71 \, h^{-1}\text{Mpc}$. This quantity is expected to be smaller than the location of the sharpest slope in three-dimensional distance from the cluster \cite{More2016}.

We also calculated the 3-dimensional splashback radius $r_{\rm sp}$ of our galaxy cluster using our model (equation \ref{eq:xi3d}). The bottom left panel of Figure \ref{fig:splashback} shows the three-dimensional cross-correlation function $\xi_{\rm 3D}(r)$, which is calculated using the posteriors of the model fit to the measurement of $\xi_{\rm 2D}(R)$. The corresponding logarithmic slope $d \log(\xi_{\rm 3D}) /d  \log(r)$ is shown in the bottom right panel. We obtained the value of the three-dimensional splashback radius $r_{\rm sp}= 2.19^{+0.50}_{-0.43} \, h^{-1}\text{Mpc}$ which is shown by the vertical blue line in the bottom two graphs. We note that the median values of $R_{\rm sp}$ and $r_{\rm sp}$ differ by roughly 18 per cent as expected in cluster scale halos (see, e.g.,\cite{More2016}).

The value of the slope $\frac{d \text{log} \xi_{\rm 3D}}{d \text{log} r}$ at the radial separation of $r=r_{\rm sp}$ is $-3.98^{+0.82}_{-1.96}$. These values are consistent with values of other X-ray selected cluster studies \cite{Divya2023, Contigiani2019} and other literature values \cite{More2016, Umetsu2017, Baxter2017, Chang2018, zuricher2019, Shin2019, Murata2020, Bianconi2021, Shin2021, Giocoli2024, Xu2024}.  %This shows that the location of the minima corresponds to the splashback feature for our cluster sample associated with the steepest slope. 

We also compared our results with the $\Lambda$CDM predictions. Using our weak lensing calibrated halo mass value of $M_{\rm 200m} = 10^{14.68} h^{-1}\text{M}_{\odot}$ at the median redshift of 0.40, we compute the value of peak height $\nu_{200\text{m}} \equiv \delta_c / \sigma(M_{200\text{m}}) / D(z)$, where $\sigma(M_{\rm 200m})$ is the variance of the linearly extrapolated overdensity field when smoothed in a spherically averaged manner within the radius $r_{\rm 200m}$. Substituting our values of $\nu_{200\text{m}}$ and $r_{200\text{m}}$ in equation 7 of \citet{More2015}, 
\begin{equation}
    \frac{r_{\rm sp}}{r_{200\rm m}} = 0.81 \left(1 + 0.97 e^{-\nu_{200\text{m}} / 2.44} \right)
\end{equation}
we obtain a predicted value of $r_{\rm sp} = 1.98 \pm 0.06 \, h^{-1}\text{Mpc}$. This is shown as the solid black line in figure \ref{fig:splashback}. Our estimated value agrees with the expected value within $1 \sigma$, indicating good agreement with $\Lambda$CDM predictions. 

\renewcommand{\arraystretch}{1.2}
\begin{table}[h]
    \centering
    \begin{tabular}{ll}
        \toprule
        \multicolumn{2}{c}{\textbf{Cluster Galaxy cross-correlations}} \\
        \midrule
		\textbf{Parameters} & \textbf{Constraints} \\ 
		\midrule
		$\log(\rho_s)$ & $2.05^{+0.49}_{-0.82}$ \\ 
		$\log(\alpha)$ & $-0.89^{+0.57}_{-0.40}$ \\ 
		$\log(r_{\rm s})$ & $-0.17^{+0.44}_{-0.29}$ \\ 
		$\log(\rho_0)$ & $0.46^{+0.27}_{-0.75}$ \\ 
		$s_{\rm e}$ & $1.09^{+0.70}_{-0.69}$ \\ 
		$\log(r_{\rm t})$ & $0.26\pm 0.12$ \\ 
		$\log(\beta)$ & $0.75^{+0.18}_{-0.19}$ \\ 
		$\log(\gamma)$ & $0.58^{+0.20}_{-0.19}$ \\ 
		$R_{\rm sp}$ & $1.71^{+0.41}_{-0.39}$ \\ 
		$r_{\rm sp}$ & $2.19^{+0.50}_{-0.43}$ \\
        $\frac{d \log \xi_{\rm 3D}}{d \log r} \mid_{r=r_{\rm sp}}$  &  $-3.98^{+0.82}_{-0.96} $\\
        $\chi_{\rm sp}^{2} / \text{dof}_{\rm eff}$ & $1.41/3.87$\\
        \\
		\bottomrule
        \multicolumn{2}{c}{\textbf{Weak Lensing Profile}} \\
        \midrule
        \textbf{Parameters} & \textbf{Constraints} \\ 
		\midrule
		$\log\ [M_{\rm 200m}/h^{-1} M_{\odot}]$ & $14.68^{+0.04}_{-0.04}$ \\ 
		$c$ & $3.48^{+0.64}_{-0.55}$ \\ 
        $r_{\rm 200m}/h^{-1} Mpc$ & $1.97^{+0.06}_{-0.06}$ \\ 
        $\chi_{\rm wl}^{2} / \text{dof}_{\rm eff}$ & $2.92/5.43$ \\
		\bottomrule
    \end{tabular}
    \caption{\textit{Posterior distribution of parameters given our measurements:} The table summarizes the median and 16th and 84th percentile uncertainties of the inferred posterior of the model parameters. Rows 1–8 list the DK14 model constraints from the galaxy number density profile. Rows 9–11 provide the two- and three-dimensional splashback radius estimates ($R_{\rm sp}$, $r_{\rm sp}$) and the 3D logarithmic slope at $r_{\rm sp}$. Rows 12–14 show the weak lensing parameters and corresponding $r_{\rm 200m}$. The final two rows report the best-fit $\chi^2$ values for the weak lensing $\chi^{2}_\mathrm{wl}$ and galaxy number density $\chi^{2}_\mathrm{sp}$ fits, along with the effective degrees of freedom from Equation 29 of \citet{Raveri2019}.}%The table shows the median values of the posterior distribution of our inferred model parameters with errors based on the 16th and 84th percentile values. The first eight rows present the parameter constraints for the DK14 model fit to our galaxy number density profile. The two-dimensional $R_{\rm sp}$ and the three-dimensional $r_{\rm sp}$ splashback radius estimates, along with the three-dimensional logarithmic slope at $r_{\rm sp}$, are mentioned in rows 9, 10 and 11, respectively. Row 12-14 presents the parameter values from the weak lensing analysis and the corresponding spherical overdensity size $r_{\mathrm{200m}}$. The last two rows present the best fit $\chi^2$ values for the weak lensing $\chi^{2}_\mathrm{wl}$ and galaxy number density profile $\chi^{2}_\mathrm{sp}$ with the associated effective degrees of freedom $\mathrm{dof}_{err}$ based on equation 29 of \citet{Raveri2019}.
    \label{tab:model_constraints}
\end{table}

%\begin{comment}
%\end{comment}

%%%%%%%%%%%%%%%%%%%%%%%%%%%%%%%%%%%%%%%%%%%%%%%%%%%%%%%%%%%%%%%%%%%%%%%%%%%%%%%%%%%%%%%%%%%%%%%%%%%%%%%%%%%%%%%%%

\subsection{\label{sec:dynamical friction}Effect of Dynamical Friction.}
\begin{figure}[hbtp]
\centering
\includegraphics[width=\linewidth]{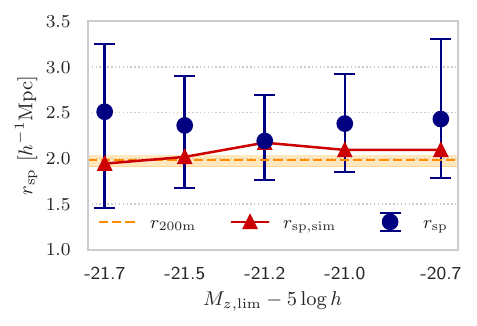}
\caption{\textit{Impact of Dynamical Friction}: The values of the three-dimensional splashback radius $r_{\rm sp}$ are shown as a function of the $z$-band absolute magnitude limit cut used for the DES Y3 optical galaxy catalog. The $r_{\rm sp}$ were calculated by modelling the DK14 profile on the 2D cross-correlation $\xi_{\rm 2D}(R)$ between the RASS-MCMF X-ray galaxy clusters and by varying the z-band magnitude limits of the DES Y3 optical galaxies. The red triangles are the splashback radius value inferred from sub halo abundance matching in MDPL2 simulation. Within the error bars, the splashback radius $r_{\rm sp}$ values are consistent with each other and with simulation values.}
\label{fig:maglim}
\end{figure}

The splashback radius calculated from galaxies may differ from that of dark matter because of dynamical friction acting on the subhalos that host the galaxies \cite{Adhikari2016}. The dynamical friction is dependent upon the ratio of the subhalo mass to that of the galaxy cluster. The larger this value, the more prominent the effect. Given that we expect more massive subhalos to host more luminous galaxies (see e.g., \cite{Dvornik2020, Wang2024, Kumar2024}), we empirically test for dynamical friction effects by cross-correlating galaxies brighter than different magnitude limits. In Figure \ref{fig:maglim}, blue points with errorbars represent the estimated value of splashback radius for each of the $z$-band magnitude limits. These values of splashback radius are consistent with each other within the error bars. 

To estimate the expected impact of dynamical friction, we make use of halo and subhalo catalogs from the MultiDark-Planck II (MDPL2), a $3840^3$ particle cosmological $N$-body simulation with a box size of $1h^{-1}\mathrm{Gpc}$ and mass resolution of $1.51 \times 10^9 h^{-11}\mathrm{M_{\odot}}$ \cite{mdpl2}. We use the $z=0.396$ particle snapshot of the MDPL2 simulation which is closest to the median redshift $z=0.4$ of our galaxy cluster sample. We use all halos identified by the six-dimensional phase space halo finder ROCKSTAR \cite{Behroozi2013} in the $z = 0.369$ snapshot with halo mass, $\mathrm{M_{200m}}$, above $3.28 \times 10^{14}$ as our sample of galaxy clusters. This galaxy cluster sample has a mean halo mass close to our fiducial sample. Given that the simulation does not have galaxy luminosities assigned to subhalos, we use subhalo abundance matching in order to select subhalos that would host the photometric galaxies we use for the measurements of the cross-correlations. We then measure the steepening of the three-dimensional density profile for both matter and subhalos that host these galaxies.

Making use of the evolving Schechter function parameters for the $z$-band magnitude from \cite{Loveday_luminosity_function}, we calculate the cumulative abundances of our photometric galaxies for various magnitude limits. We first obtain an estimate of the $V_{\rm peak}$ of subhalos hosting our galaxies by matching this cumulative abundance of our photometric galaxies with that of the dark matter subhalos in the MDPL2 simulation (See Appendix \ref{app:sham}). Note that, we do not include any scatter between $V_{\rm peak}$ of subhalos and the magnitude of galaxies. This is a conservative approach since it will give the largest possible impact due to dynamical friction \cite{More2016}. The subhalos that host our fiducial subsample of photometric galaxies approximately correspond to subhalos with $V_{peak}>190.16 \mathrm{\,km/s}$ while the brightest and faintest subsamples correspond to $V_{peak}>242.98 \mathrm{\,km/s}$ and $V_{peak}>157.74 \mathrm{\,km/s}$ respectively.

\begin{figure}[hbtp]
\centering
\includegraphics[width=\linewidth]{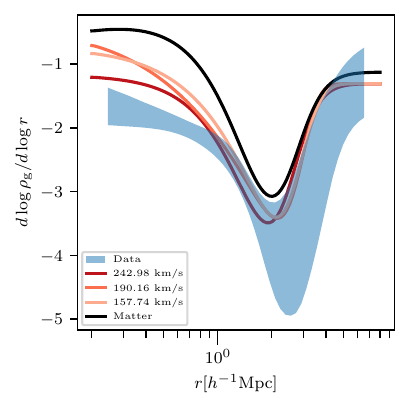}
\caption{\textit{SHAM}: Comparison between the logarithmic slope of the density profile for matter and that of subhalos selected using different $V_{peak}$ thresholds from the MDPL2 simulation data as indicated in the legend. The blue shaded region is our fiducial estimate of our logarithmic slope. }
\label{fig:sham_rsp}
\end{figure}

We calculate the 3-dimensional number density profile of subhalos greater than our calculated $V_{\rm peak}$ values. In Figure \ref{fig:sham_rsp}, we show the logarithmic slopes of subhalo distributions around galaxy cluster halos for different $V_{\rm peak}$ thresholds obtained from our simple subhalo abundance matching method. The blue shaded region is our estimate of logarithmic slopes for our fiducial sample. We observe that the location of the steepest slope for subhalos with the lowest $V_{peak}$ is similar to that of dark matter within $5\%$.

The expected values of the splashback radius and their dependence on the magnitude limit are shown as red triangular data points in Figure~\ref{fig:maglim}. We can see that the variation in the values of the splashback radius for our subsamples is not expected to be large and is entirely consistent with our observations given the statistical errors.

Our results from simulations indicate that the steepening of the three-dimensional slopes for both matter and subhalos that host our ﬁducial photometric sample of galaxies are expected to occur at similar location. This is consistent with \citet{Oshea2024}, who used the IllustrisTNG cosmological simulation and showed that dynamical friction does not have a significant effect on clusters with $M_{200,\mathrm{mean}}>10^{14} \mathrm{M_{\odot}}$ when considering subhalos that host galaxies as massive as $10^{10}M_\odot$. Thus, dynamical friction does not seem to have significant impact on our inferred location of splashback radius.

%We measured the cross-correlation signal for three more $z$-band absolute magnitude limits of -21.7, -21.5 and -21.2 for the DES Y3 galaxies. The value of $r_{\rm sp}$ remains reasonably consistent within the errors. For brighter magnitudes, the cross-correlation signal had huge uncertainties. Figure \ref{fig:maglim} shows the measured value of the three-dimensional splashback radius $R_{\rm sp}$ for different values of $i$-band magnitude limits. This shows that dynamical friction does not create any systematics in the measurement of our splashback radius estimate. This is consistent with \citet{Oshea2024}, which showed through a simulation study that for massive clusters, the effect of dynamical friction is minimal. 

\subsection{\label{sec:comparisons}Comparison with literature}

\begin{figure*}[hbtp!]
\centering
\includegraphics[width=\textwidth]{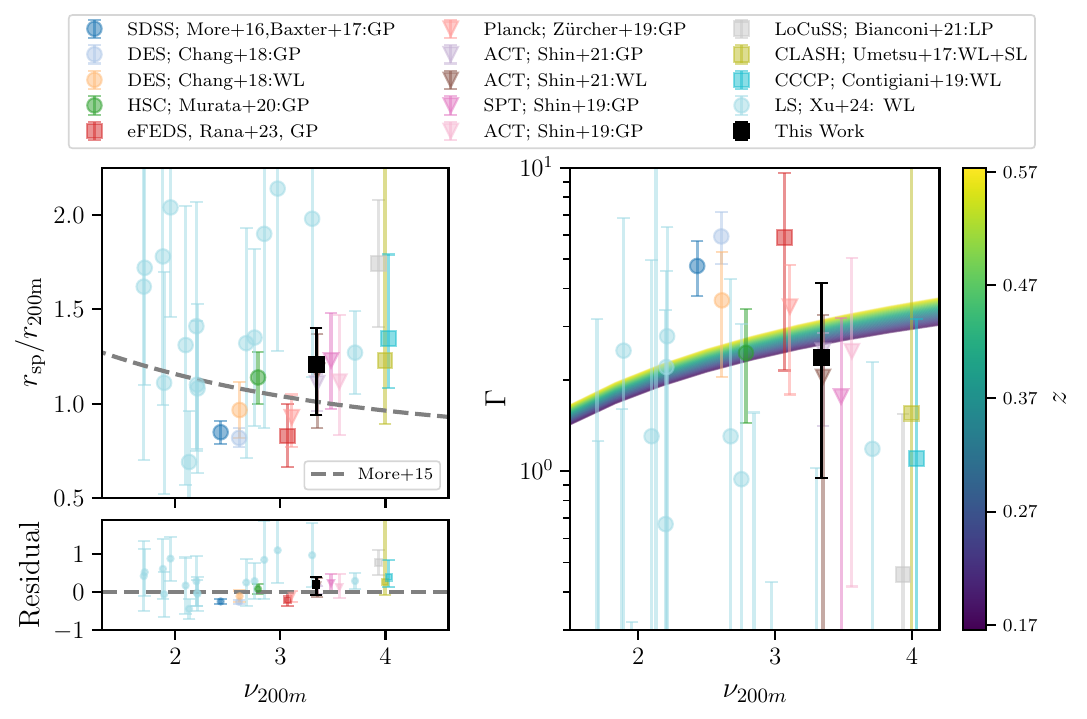}
\caption{\textit{$r_{\rm sp}$ Literature comparison}: In this figure, we compare our splashback radius results with other existing results in the literature. In the top right panel, thex-axis is the peak height $\nu_{200m}$ while the y-axis represents the ratio of splashback radius $r_{\rm sp}$ to the spherical overdensity $r_{\rm 200m}$ radius. The grey dashed line illustrates the expected behaviour from $\Lambda$CDM using the fitting function provided by \citet{More2015}. The right bottom figure is the residual difference between the data points and the grey line. In the right hand panel, we show the mean accretion rate $\Gamma$ as a function of the peak height. Colored region shows expectations from \citet{Diemer2017}, and the colorbar indicates redshifts from sample minimum to maximum. In all the graphs, the circular data points represent values calculated from optically selected clusters, the triangular points correspond to results from SZ-selected clusters, and the square points are based on X-ray selected clusters. In the legend, we label each set of data points according to their survey name, reference, and the analysis method used — weak lensing (WL), strong lensing (SL), galaxy number density (GP), or stacked luminosity profile (LP).  The black data point, with error bars, represents our findings from this study using the RASS-MCMF X-ray clusters.}
\label{fig:vpeak_graphs}
\end{figure*}

In this section, we compare our results with previous studies in the literature. In Figure \ref{fig:vpeak_graphs}, we compare our findings with those from the literature in the $\nu_{\rm 200m}$--$r_{\rm sp}/r_{\rm 200m}$ plane, where $\nu_{\rm 200m}$ is the peak height of the cluster halos used to measure the splashback radius. Since this relation is largely independent of redshift \cite{More2015}, it facilitates a consistent comparison across various studies which have used clusters of different masses and at various median redshifts. The grey dashed line illustrates the expectations from $\Lambda$CDM, using the parameterization provided by eqn. 7 in \citet{More2015}, which was calibrated using numerical simulations. The circular data points represent constraints from optically selected clusters, triangular data points indicate results from clusters identified via the SZ effect, and square data points correspond to X-ray selected clusters. In the legend, we have labelled the dataset along with the method used in each study—galaxy number density profile (GP), weak lensing (WL), and stacked luminosity profile (LP). The black square point marks our constraint derived from the galaxy number density profile around the RASS-MCMF X-ray clusters from this work, yielding $r_{\rm sp}/r_{\rm 200m} \approx 1.11^{+0.26}_{-0.22}$. This result represents a 20 per cent constraint and is consistent with the $\Lambda$CDM prediction within $\approx 1 \sigma$. Our finding has similar accuracy compared to the cross-correlation analysis conducted on the eFEDS X-ray cluster sample by \citet{Divya2023}, which reported a value of $r_{\rm sp}/r_{\rm 200m} \approx 0.83 \pm 0.17$ but at a lower peak height, i.e., cluster masses. Similarly, our analysis puts a tighter bounds when compared with CLASH X-ray cluster sample by \citet{Umetsu2017} and CCCP X-ray cluster study by \citet{Contigiani2019}.

The splashback radius compared to the traditional boundary radius $r_{\rm 200m}$ depends upon the the accretion rate of dark matter halos defined by equation 3 of \citet{Diemer_Kravtsov_2014},
\begin{equation}
    \Gamma = \Delta \log(M_{\mathrm{vir}})/\Delta \log(a)\,.
\end{equation}
We infer the mean accretion rate of galaxy clusters by substituting the median redshift of $\langle z \rangle$ of our cluster sample  and the value of our inferred value of $r_{\rm sp}/r_{\rm 200m}$ into equation 5 of \citet{More2015}
\begin{equation}
    \Gamma = 0.916 - 3.04 \; \mathrm{ln}\left( \frac{r_{\rm sp}/r_\mathrm{200m}}{0.54+0.29 \; \Omega (\langle z \rangle)}-1 \right)
\end{equation}
A similar relation for the median accretion rate is provided by equation 5 of \citet{More2016}. The difference between the mean and median accretion rate values is small compared to the size of the error bars, and thus does not significantly affect our interpretation.

The right side graph in Figure \ref{fig:vpeak_graphs} shows the $\Gamma$--$\nu_{200m}$ graph for different literature values. The colored shaded region represents the expectations from \citet{Diemer2017}, with the colorbar indicating the corresponding redshifts, spanning from the minimum to the maximum redshift of the literature samples. The results from our study agree with the expected values within $1\sigma$ although with large errorbars. We note however that this is the first time estimate of the accretion rates of galaxy clusters have been inferred using the location of the inferred splashback radius. Our X-ray analysis provides better constraints than previous X-ray studies and yield results which have accuracy comparable to those derived from SZ cluster studies.

% \begin{figure}[hbtp]
% \centering
% \includegraphics[width=\linewidth]{plots_pdf/accretion_rate_vpeak_no_legend.pdf}
% \caption{\textit{$\Gamma$ Comparison with other studies}: We compare our accretion rate values with other existing results in the literature. The x-axis is the peak height $v_{200m}$ while the y-axis represents the mass accretion rate $\Gamma$ as in \citet{Diemer_Kravtsov_2014}. The circular data points represent values calculated from optically selected clusters. In contrast, the triangular points correspond to results from SZ-selected clusters, and the square points are based on X-ray selected clusters. In the legend, we label each set of data points according to their survey name, reference, and the analysis method used — weak lensing (WL), strong lensing (SL), galaxy number density (GP), or stacked luminosity profile (LP). The brown shaded contour indicates the uncertainty on the mean of expected values and the dashed brown lines represent the 68$\%$ interval around it as in Figure 8 of \citet{Diemer_Kravtsov_2014}.}
% \label{fig:accretion_peak}
% \end{figure}

\section{\label{sec:conclusion} Conclusion}
The splashback radius is a physically motivated definition of the halo boundary as compared to traditional definitions, which are dependent on redshift and the accretion rate of the halo. Optically selected galaxy clusters still provide the best determinations of the location of the splashback radius, although numerous studies have questioned whether optical selection effects could systematically bias these measurements. We therefore use X-ray selected galaxy clusters from the RASS-MCMF catalog for the purpose of splashback radius measurements. We use a luminosity threshold $L_{\rm X} = 10^{44} \text{ergs} \,\text{h}^{-2} \text{s}^{-1}$ and redshift $z<0.6$ to obtain an approximately volume-limited sample of 255 galaxy clusters. Making use of the DES Y3 RedMaGic galaxy catalog we calculate the cluster-galaxy cross-correlation signal to infer the location of the splashback radius. The halo mass $M_{\rm 200m}$ and the corresponding traditional halo radius $r_{\rm 200m}$ was calculated from weak lensing analysis using the DES Y3 shape catalog. 

The findings of our analysis are summarized below.

\begin{itemize}
\item Using the DES Y3 shape catalog, we calculate the stacked weak lensing signal of our cluster sample with a signal-to-noise ratio of 32.19. We use a simple NFW profile to model this signal and found the mean of the cluster halo mass to be $\log (M_{\rm 200m} / h^{-1} M_\odot) = 14.68_{-0.04}^{+0.04}$. Using this, we calculate the traditional halo boundary $r_{\rm 200m} = 1.97_{-0.06}^{+0.05}\, h^{-1} \text{Mpc}$ of our cluster sample. 

\item We also model the weak lensing signal using the DK14 profile \cite{Diemer_Kravtsov_2014}. Although we were not able to get any statistically significant constraints on the location of the splashback radius from weak lensing alone,  we explicitly showed that $\log (M_{\rm 200m} / h^{-1} M_\odot) = 14.66_{-0.03}^{+0.03}$, a value consistent with the cluster halo mass assuming the NFW profile. 

\item We measure the projected cross-correlation signal of our galaxy clusters with a signal-to-noise ratio of 16.41 using the DES Y3 RedMaGiC galaxies having $z$-band absolute magnitude cut $M_z - 5 \log h < -21.2$. We model this measurement using the DK14 profile \cite{Diemer_Kravtsov_2014} to infer the location of the steepest slope and obtain the splashback radius measurement. 

\item The value of this steepest slope was found to be $-3.97^{+0.82}_{-0.96}$, which is steeper than the  expectation from the NFW profile, especially when the two halo term is considered. This provides evidence for the presence of steepening associated with the splashback radius.

\item We infer the location of the projected splashback feature $R_{sp}$/$h^{-1} \text{Mpc} = 1.71^{+0.41}_{-0.39}$ with $\approx 22$ percent constraint and the three-dimensional splashback value of $r_{sp}$/$h^{-1} \text{Mpc} = 2.19^{+0.50}_{-0.45}$ with $\approx 18$ percent constraint. These values are in the range we expect for the massive halos in numerical simulations \cite{Diemer_Kravtsov_2014, More2015}.

\item These constraints on the three-dimension splashback radius $r_{sp}$ are comparable to the value from the spherical overdensity estimates $r_{\rm 200m} = 1.97_{-0.06}^{+0.05} h^{-1} \text{Mpc}$. These values are consistent ($\approx 1 \sigma$) with the expectations from numerical simulations for the weak lensing calibrated halo mass at the median redshift of $z = 0.40$. These results have improved constraints as compared to previous X-ray selected galaxy cluster studies such as \citet{Divya2023}.

\item We calculate the halo mass accretion rate $\Gamma = 2.38^{+1.78}_{-1.43}$ using equation 5 of \citet{More2015}. The large errors are dominated by the statistical errors on the inferred splashback radius $r_{sp}$.

\end{itemize}

In our analysis, we are only using the $z$-band magnitude of DES Y3 RedMaGiC galaxy catalog for our cross-correlation measurements. We have calculated the value of splashback radius using $z$-band absolute magnitude limits of -21.7, -21.5, -21.2, -21.0 and -20.7. We investigated the potential impact of dynamical friction on the inferred splashback radius by comparing the three-dimensional density profiles of dark matter and subhalos hosting galaxies, using the MDPL2 simulation.  Our results demonstrate that the location of the steepest slope in the subhalo profile closely matches that of the dark matter, particularly for our fiducial sample, with discrepancies at most within $5\%$. The measured splashback radii for galaxy samples across different magnitude thresholds are consistent within uncertainties and align well with those inferred from the subhalo-matched distributions. This shows that the effect of dynamical friction does not play a considerable role in shaping the splashback feature for our sample, a result that is also supported by recent findings using illustrisTNG simulation \cite{Oshea2024}.

Our results provide improved constraints on the splashback radius of X-ray selected galaxy clusters than previous studies; the errors are still statistically large. However, in the near future, we expect better constraints by using all-sky cluster catalogues from the eRASS \cite{Bulbul2024}, which would help in a thorough investigation of differences between the theoretical predictions and observations. It will also allow us to explore the dependence of the splashback feature on X-ray cluster observables such as dynamical state, X-ray morphology and hydrostatic halo masses.

Finally, the accretion rate for halos of a given mass and redshift are expected to be dependent on cosmology. Measurements of the splashback radius of X-ray galaxy clusters, their masses through weak gravitational lensing and hence the mass accretion rates can thus provide an independent check of the concordance cosmological model.

\begin{acknowledgments}
 We acknowledge the use of the high performance computing facility - Pegasus at IUCAA. 

 %%%%%%%%%%%%%%%%%%%%%% DES Y3 %%%%%%%%%%%%%%%%%%%%%%%%%%%%%%%%
 This project used public archival data from the Dark Energy Survey (DES). Funding for the DES Projects has been provided by the U.S. Department of Energy, the U.S. National Science Foundation, the Ministry of Science and Education of Spain, the Science and Technology FacilitiesCouncil of the United Kingdom, the Higher Education Funding Council for England, the National Center for Supercomputing Applications at the University of Illinois at Urbana-Champaign, the Kavli Institute of Cosmological Physics at the University of Chicago, the Center for Cosmology and Astro-Particle Physics at the Ohio State University, the Mitchell Institute for Fundamental Physics and Astronomy at Texas A\&M University, Financiadora de Estudos e Projetos, Funda{\c c}{\~a}o Carlos Chagas Filho de Amparo {\`a} Pesquisa do Estado do Rio de Janeiro, Conselho Nacional de Desenvolvimento Cient{\'i}fico e Tecnol{\'o}gico and the Minist{\'e}rio da Ci{\^e}ncia, Tecnologia e Inova{\c c}{\~a}o, the Deutsche Forschungsgemeinschaft, and the Collaborating Institutions in the Dark Energy Survey.

The Collaborating Institutions are Argonne National Laboratory, the University of California at Santa Cruz, the University of Cambridge, Centro de Investigaciones Energ{\'e}ticas, Medioambientales y Tecnol{\'o}gicas-Madrid, the University of Chicago, University College London, the DES-Brazil Consortium, the University of Edinburgh, the Eidgen{\"o}ssische Technische Hochschule (ETH) Z{\"u}rich,  Fermi National Accelerator Laboratory, the University of Illinois at Urbana-Champaign, the Institut de Ci{\`e}ncies de l'Espai (IEEC/CSIC), the Institut de F{\'i}sica d'Altes Energies, Lawrence Berkeley National Laboratory, the Ludwig-Maximilians Universit{\"a}t M{\"u}nchen and the associated Excellence Cluster Universe, the University of Michigan, the National Optical Astronomy Observatory, the University of Nottingham, The Ohio State University, the OzDES Membership Consortium, the University of Pennsylvania, the University of Portsmouth, SLAC National Accelerator Laboratory, Stanford University, the University of Sussex, and Texas A\&M University.

Based in part on observations at Cerro Tololo Inter-American Observatory, National Optical Astronomy Observatory, which is operated by the Association of Universities for Research in Astronomy (AURA) under a cooperative agreement with the National Science Foundation.

 %%%%%%%%%%%%%%%%%%%%%%% MDPL2 %%%%%%%%%%%%%%%%%%%%%%%%%%%%%%%%
 The CosmoSim database used in this paper is a service by the Leibniz-Institute for Astrophysics Potsdam (AIP). The MultiDark database was developed in cooperation with the Spanish MultiDark Consolider Project CSD2009-00064. The authors gratefully acknowledge the Gauss Centre for Supercomputing e.V. (www.gauss-centre.eu) and the Partnership for Advanced Supercomputing in Europe (PRACE, www.prace-ri.eu) for funding the MultiDark simulation project by providing computing time on the GCS Supercomputer SuperMUC at Leibniz Supercomputing Centre (LRZ, www.lrz.de). The Bolshoi simulations have been performed within the Bolshoi project of the University of California High-Performance AstroComputing Center (UC-HiPACC) and were run at the NASA Ames Research Center. 
 
 DR acknowledges funding from the European Research Council (ERC) under the European Union’s Horizon 2020 research and innovation program (Grant agreement No. 101053992).
\end{acknowledgments}

% \appendix

\section{Appendixes}

% To start the appendixes, use the \verb+\appendix+ command.
% This signals that all following section commands refer to appendixes
% instead of regular sections. Therefore, the \verb+\appendix+ command
% should be used only once---to setup the section commands to act as
% appendixes. Thereafter normal section commands are used. The heading
% for a section can be left empty. For example,
% \begin{verbatim}
% \appendix
% \section{}
% \end{verbatim}
% will produce an appendix heading that says ``APPENDIX A'' and
%\begin{verbatim}
\appendix

%%%%%%%%%%%%%%%%%%%%%%%%%%%%%%%%%%%%%%%%%%%%%%%%%%%%%%%%%%%%%%%%%%%%%%%%%%%%%%%%%
\renewcommand{\thefigure}{\Alph{section}\arabic{figure}}
\setcounter{figure}{0} % Reset figure counter
\section{Boost Parameters}\label{App:boost parameter}
\begin{figure}[hbtp!]
    \centering
    \includegraphics[width=0.9\linewidth]{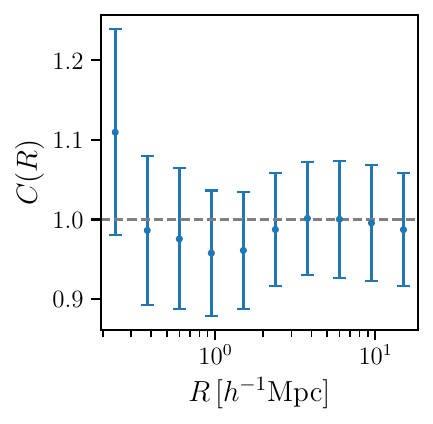}
    \caption{\textit{Boost parameters:} The data points show the boost parameter $C(R)$ for our weak lensing measurement $\Delta\Sigma$ estimated using equation \ref{eq:boost}. The plot shows consistency of boost parameter with grey dashed unity line}
    \label{fig:boost_parameter}
\end{figure}

In Figure \ref{fig:boost_parameter}, we present our boost parameter $C(R)$ measurements. The blue datapoints represents the stacked measurement calculated by Eq. \ref{eq:boost} using our cluster sample and 20 times larger randoms. The corresponding errors are computed using 20 area jacknife realizations. Since the measurements are consistent with unity within statistical uncertainties, we do not apply any boost correction in this work.

%%%%%%%%%%%%%%%%%%%%%%%%%%%%%%%%%%%%%%%%%%%%%%%%%%%%%%%%%%%%%%%%%%%%%%%%%%%%%%%%%
\section{Covariances}\label{App:covariance}
Figure \ref{fig:wl_corr} and Figure \ref{fig:splash_corr} show the correlation coefficient $r_{ij}$ for the corresponding covariances of the weak lensing and galaxy density profile measurements. Given the covariance $C$, we calculate the correlation coefficient given by
\begin{equation}
    r_{\rm ij} = \frac{C_{\rm ij}}{\sqrt{C_{\rm ii}C_{\rm jj}}}
\end{equation}
where subscripts ij represent the $\text{i}^{\rm th}$ and $\text{j}^{\rm th}$ radial bins. We use the jackknifes to compute the covariances for the galaxy number density profile and shape noise covariance for the weak lensing profile. We describe the details in Section \ref{sec:model}.\\
\setcounter{figure}{0}
\begin{figure}[hbtp!]
    \centering
    \includegraphics[width=0.9\linewidth]{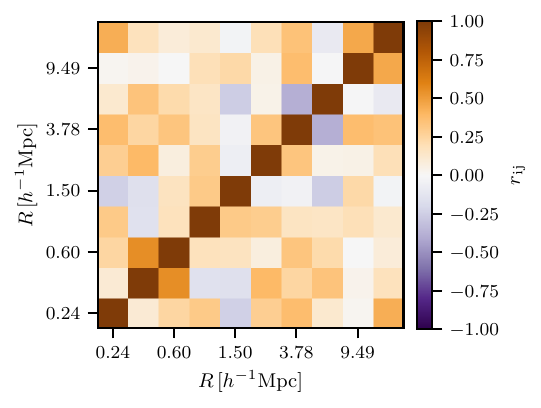}
    \caption{\textit{Weak lensing profile covariance:} The above plot shows the correlation coefficient $r_{ij}$ matrix for the $i^{\mathrm{th}}$ and $j^{\mathrm{th}}$ radial bin of the weak lensing signal measurements $\Delta \Sigma$ given in Section \ref{sec:wl_model} and estimated by shape noise using 200 different random rotations of background source galaxies with x-y axis showing our radial binning for signal measurements.}
    \label{fig:wl_corr}
\end{figure}
\begin{figure}[h!]
    \centering
    \includegraphics[width=0.9\linewidth]{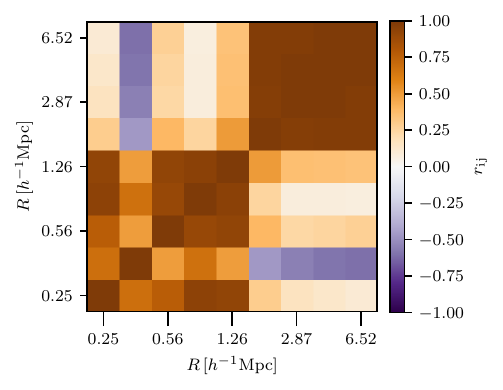}
    \caption{\textit{Galaxy number density covariances:} The above plot shows the correlation coefficient $r_{ij}$ matrix for the $i^{\mathrm{th}}$ and $j^{\mathrm{th}}$ radial bin of the galaxy number density profile measurement $\xi_{\mathrm{2D}}$ given in Section \ref{sec:dk14} and estimated using 20 jackknife regions with x-y axis showing our radial binning for signal measurements.}
    \label{fig:splash_corr}
\end{figure}\\
Figure \ref{fig:wl_corr} shows the weak lensing correlations. We see no correlation among different radial bin signals, as it is dominated by the shape noise at the scales of our interest. The source galaxies that contribute to the signal in each radial bin are different and thus independent. At the same time, we found a mild positive correlation at small scales ($R \leq 1.26 h^{-1} {\mathrm Mpc}$) for the galaxy number density measurement shown by the off-diagonal elements in Figure \ref{fig:splash_corr}. This is consistent with correlations in the small scale clustering of satellite galaxies around clusters. At larger radii ($R > 1.26 h^{-1} {\mathrm Mpc}$), the correlations are mostly weak, with some outer bins showing a slight negative correlation with the inner bins. This could reflect small differences in the satellite distribution across clusters.  %\todoin{Write about the number density correlation profile}

%%%%%%%%%%%%%%%%%%%%%%%%%%%%%%%%%%%%%%%%%%%%%%%%%%%%%%%%%%%%%%%%%%%%%%%%%%%%%%%%%
\section{Mass comparison}
\setcounter{figure}{0}
\begin{figure}[h!]
    \centering
    \includegraphics[width=0.9\linewidth]{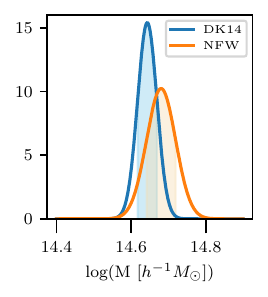}
    \caption{\textit{Mass comparison:} The posterior distribution of Mass $m_{\mathrm{200m}}$ inferred based on the DK14 profile model (blue) comparedto the NFW (orange) profile model when fit to the weak lensing signal measurements.}
    \label{fig:dk14_nfw_comp}
\end{figure}
Figure \ref{fig:dk14_nfw_comp} shows the posterior distribution of mass $M_{200m}$ calculated using the DK14 as well as NFW fit to the weak lensing profile as discussed section \ref{sec:weak_lensing}. We can see that these values agree with each well within $1 \sigma$. 
%%%%%%%%%%%%%%%%%%%%%%%%%%%%%%%%%%%%%%%%%%%%%%%%%%%%%%%%%%%%%%%%%%%%%%%%%%%%%%%%%
\section{\label{app:sham}Sub Halo Abundance Matching}
We use a simple subhalo abundance matching technique to determine the properties of subhalos hosting the photometric galaxies we use in this paper. The left-hand side of Figure \ref{fig:sham_appendix} shows the cumulative abundance of galaxies, based on Schechter function fit to $z$-band luminosity function obtained from \citet{Loveday_luminosity_function}. The magnitude limits for our photometric sub-samples are shown in red dotted vertical lines, while the horizontal dotted lines show the cumulative abundances. We match these abundances to the observed abundances as a function of $V_{peak}$. In the right-hand panel of Figure \ref{fig:sham_appendix}, the dotted vertical lines show these matched $V_{peak}$ threshold values. 

Using the subhalo abundance matching, the subhalos hosting our fiducial subsample of galaxies having $z$-band magnitude limit of $-21.2$ correspond to subhalos with $V_{peak}>190 \mathrm{km/s}$. We perform abundance matching for brighter and fainter magnitude limit subsamples shown in Figure \ref{fig:maglim}. We also perform abundance matching for one more brighter subsample corresponding to $V_{peak}>290 \mathrm{km/s}$. In section \ref{sec:dynamical friction}, we explore how dynamical friction is expected to affect the location of splashback radius.  
\setcounter{figure}{0}
\begin{figure*}%[hbtp!]
    \centering
    \includegraphics[width=\textwidth]{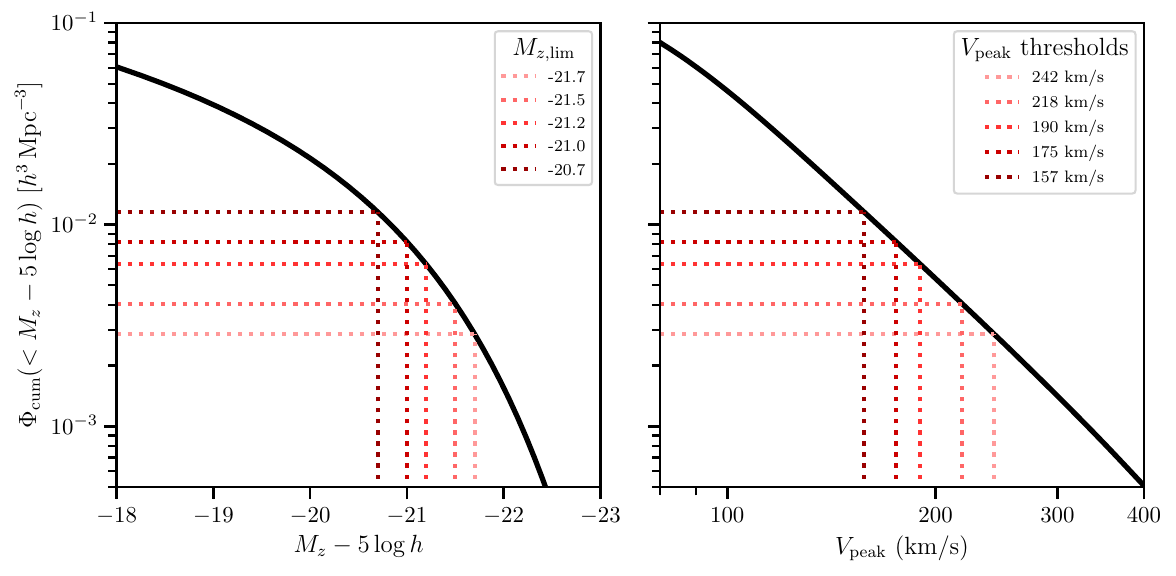}
    \caption{\textit{Left:} the cumulative abundance of GAMA galaxies as a function of their $z$-band magnitude plotted using evolving schechter function parameters from \citet{Loveday_luminosity_function} at the median redshift of our sample, $z=0.4$. The dotted vertical lines correspond to the magnitude limits of our photometric samples of galaxies. The horizontal lines give the cumulative abundance. \textit{Right:} the cumulative abundances of subhalos in MDPL2 simulation as a function of $V_{\rm peak}$. The horizontal dotted lines are the cumulative abundances of our photometric samples taken from the left panel, while the vertical dashed lines give the $V_{\rm peak}$ thresholds corresponding to our photometric subsamples.}
    \label{fig:sham_appendix}
\end{figure*}

%%%%%%%%%%%%%%%%%%%%%%%%%%%%%%%%%%%%%%%%%%%%%%%%%%%%%%%%%%%%%%%%%%%%%%%%%%%%%%%%%
% \newpage
\section{Degeneracies in the Model Parameters}
Figure \ref{fig:splash_corr} shows the correlations between the parameters used in the DK14 model fit to the cluster-galaxy cross-correlation measurements. The functional form presented in equations \ref{eq:rhoin}, \ref{eq:rhout}, and \ref{eq:ftrans} are the main reasons for the parametric degeneracies. We see that all the posteriors of the parameters are well-behaved and do not pile-up against the boundaries of the adopted priors. 

\setcounter{figure}{0}
\begin{figure*}[h!]%[p]
\centering
\includegraphics[width=\textwidth]{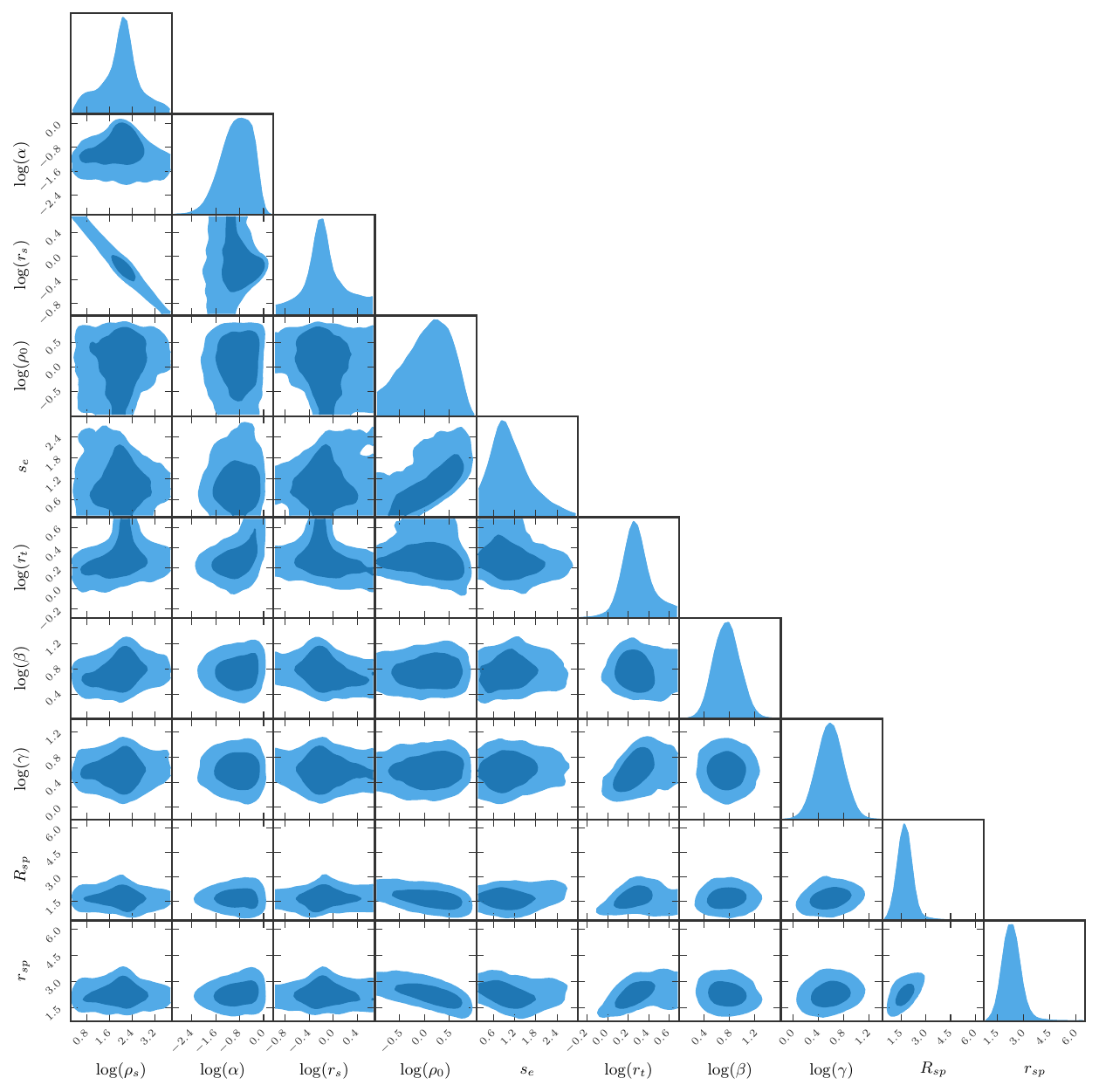}
    \caption{\textit{Splashback parameter posteriors:} This corner plot shows the posterior of the model parameters and their degeneracies when fit to the projected cluster-galaxy cross-correlation measurements.}
    \label{fig:splash_contour}
\end{figure*}

%\end{verbatim}
% will produce an appendix heading that says ``APPENDIX A: BACKGROUND''
% (note that the colon is set automatically).

% If there is only one appendix, then the letter ``A'' should not
% appear. This is suppressed by using the star version of the appendix
% command (\verb+\appendix*+ in the place of \verb+\appendix+).

% \section{A little more on appendixes}

% Observe that this appendix was started by using
% \begin{verbatim}
% \section{A little more on appendixes}
% \end{verbatim}

% Note the equation number in an appendix:
% \begin{equation}
% E=mc^2.
% \end{equation}

% \subsection{\label{app:subsec}A subsection in an appendix}

% You can use a subsection or subsubsection in an appendix. Note the
% numbering: we are now in Appendix~\ref{app:subsec}.

% Note the equation numbers in this appendix, produced with the
% subequations environment:
% \begin{subequations}
% \begin{eqnarray}
% E&=&mc, \label{appa}
% \\
% E&=&mc^2, \label{appb}
% \\
% E&\agt& mc^3. \label{appc}
% \end{eqnarray}
% \end{subequations}
% They turn out to be Eqs.~(\ref{appa}), (\ref{appb}), and (\ref{appc}).

\nocite{*}

% \bibliography{bibliography}% Produces the bibliography via BibTeX.
\providecommand{\noopsort}[1]{}\providecommand{\singleletter}[1]{#1}%

\end{document}